\documentclass[a4paper,11pt]{article}
\usepackage{setspace}
\usepackage[left = 2cm, right=2cm, top=3cm, bottom=3cm]{geometry}
\usepackage[pdftex]{graphicx}
\usepackage{cite}
\usepackage{bm}
\usepackage{float}
\usepackage{amsmath,amssymb,latexsym}
\usepackage{color}
\usepackage[T1]{fontenc}
\usepackage{wasysym}
\usepackage{siunitx}
\usepackage{setspace}
\usepackage{relsize}
\usepackage{tikz}
\usepackage{lineno, blindtext}
\usepackage{authblk}
\usepackage[symbol]{footmisc}
\usepackage{lineno}
\usepackage[pdfencoding=auto, psdextra]{hyperref}

\begin{document}

	\renewcommand{\figurename}{Fig.}
	
	\title{\color{blue}\textbf{Using optical tweezer electrophoresis to investigate clay nanoplatelet adsorption on Latex microspheres in aqueous media}}
	\author[1]{Vaibhav Raj Singh Parmar}
	\affil[1]{\textit{Soft Condensed Matter Group, Raman Research Institute, C. V. Raman Avenue, Sadashivanagar, Bangalore 560 080, INDIA}}
	\author[1]{Sayantan Chanda}
        \author[1]{Sri Vishnu Bharat Sivasubramaniam}
	\author[1,*]{Ranjini Bandyopadhyay}
	
	\footnotetext[1]{Corresponding Author: Ranjini Bandyopadhyay; Email: ranjini@rri.res.in}
	\maketitle
	\begin{abstract}
		The adsorption of charged clay nanoplatelets plays an important role in stabilizing emulsions by forming a barrier around the emulsion droplets and preventing coalescence. In this work, the adsorption of charged clay nanoplatelets on a preformed Latex microsphere in an aqueous medium is investigated at high temporal resolution using optical tweezer-based single-colloid electrophoresis. Above a critical clay concentration, charged clay nanoplatelets in an aqueous medium self-assemble gradually to form gel-like networks that become denser with increasing medium salinity. In a previous publication [R. Biswas \textit{et. al.}, \textbf{Soft Matter}, 2023, \textbf{19}, 24007-2416], some of us had demonstrated that a Latex microsphere, optically trapped in a clay gel medium, is expected to attach to the network strands of the gel. In the present contribution, we show that for different ionic conditions of the suspending medium, the adsorption of clay nanoplatelets increases the effective surface charge on an optically trapped Latex microsphere while also enhancing the drag experienced by the latter. Besides the ubiquitous contribution of non-electrostatic dispersion forces in driving the adsorption process, we demonstrate the presence of an electrostatically-driven adsorption mechanism when the microsphere was trapped in a clay gel. These observations are qualitatively verified via cryogenic field emission scanning electron microscopy and are useful in achieving colloidal stabilisation, for example, during the preparation of clay-armoured Latex particles in Pickering emulsion polymerisation.
	\end{abstract}
	\definecolor{black}{rgb}{0.0, 0.0, 0.0}
	\definecolor{red(ryb)}{rgb}{1.0, 0.15, 0.07}
	\definecolor{darkred}{rgb}{0.55, 0.0, 0.0}
	\definecolor{blue(ryb)}{rgb}{0.01, 0.2, 1.0}
	\definecolor{darkcyan}{rgb}{0.0, 0.55, 0.55}
	\definecolor{navyblue}{rgb}{0.0, 0.0, 0.5}
	\definecolor{olivedrab(web)(olivedrab3)}{rgb}{0.42, 0.56, 0.14}
	\definecolor{darkraspberry}{rgb}{0.53, 0.15, 0.34}
	\definecolor{magenta}{rgb}{1.0, 0.0, 1.0}
	
	\newcommand{\blsquare}{\textcolor{black}{\small$\blacksquare$}}
	\newcommand{\hlsquare}{\textcolor{darkred}{\small$\square$}}
	\newcommand{\redtraingle}{\textcolor{magenta}{\small$\triangle$}}
	\newcommand{\oolive}{\textcolor{olivedrab(web)(olivedrab3)}{\large$\circ$}}
	\newcommand{\purpletraingle}{\textcolor{darkraspberry}{\small$\triangledown$}}
	
	\newcommand{\bltriangle}{\textcolor{black}{\small$\triangleup$}}
	\newcommand{\rcircle}{\textcolor{red(ryb)}{\large$\bullet$}}
	\newcommand{\rtraingle}{\textcolor{red(ryb)}{\small$\triangledown$}}
	\newcommand{\wine}{\textcolor{darkred}{\large$\bullet$}}
	\newcommand{\cyan}{\textcolor{darkcyan}{\large$\bullet$}}
	\newcommand{\owine}{\textcolor{darkred}{\large$\circ$}}
	\newcommand{\redcircle}{{\large$\circ$}}
	\newcommand{\ocyan}{\textcolor{darkcyan}{\large$\circ$}}
	\newcommand{\blue}{\textcolor{blue(ryb)}{\large$\bullet$}}
	\newcommand{\blbullet}{\textcolor{navyblue}{\large$\bullet$}}
	\newcommand{\olbullet}{\textcolor{olivedrab(web)(olivedrab3)}{\large$\bullet$}}
	\newcommand{\hollowblue}{\textcolor{blue(ryb)}{\large$\circ$}}
	\newcommand{\black}{\textcolor{black}{\large$\bullet$}}
	\newcommand{\hollowblack}{\textcolor{black}{\large$\circ$}}
	\newcommand{\bltria}{\textcolor{black}{\small$\triangle$}}
	\newcommand{\rtrai}{\textcolor{red(ryb)}{\large$\triangledown$}}
	
	\section{Introduction}
	The attachment or adsorption of colloidal particles on solid surfaces in aqueous media decreases the overall interfacial energy of the system and is usually governed by screened electrostatic forces and non-electrostatic intermolecular forces~\cite{ZHANG200035,doi:10.1021/la502588g,doi:10.1126/science.1135752,Adeyemo2017,doi:10.1021/acs.langmuir.5b03576}. Adsorption is a ubiquitous surface phenomenon with implications in surface coating, stabilisation, pollutant removal and heterogeneous catalysis~\cite{GICHEVA201351,doi:https://doi.org/10.1002/9781118881194.ch7}.
Colloidal clay nanoplatelets are characterised by shape-anisotropy and a high surface-to-volume ratio. Clays are useful in many processes, such as in additive manufacturing, and are used as rheological modifiers in industrial products. 
{Previous} studies on surfactant-free emulsion polymerization of Laponite clay-armored Latex microspheres demonstrated that {Laponite }clay nanoplatelets adsorbed on the surface{s} of the microspheres to form monolayer{s} that transformed into multilayers with increasing clay concentration~\cite{doi:10.1021/acs.langmuir.6b01080,doi:10.1021/acs.langmuir.5b03576, Teixeira2011,B904740A}. It was concluded from these experiments that non-electrostatic attractions between the clay and Latex particles resulted in clay adsorption. 

Laponite, a synthetic hectorite clay belonging to the family of smectites, has a crystalline structure that is typical of 2:1 phyllosilicates~\cite{claybook,THOMPSON1992236}. Each particle is disk-shaped with a diameter of $25$-$30$ nm and a thickness of $0.92$ nm~\cite{claybook}. Clay nanoplatelets are arranged in one-dimensional stacks or tactoids in dry powder form and have negatively charged faces. Neighbouring nanoplatelets in a tactoid share Na$^+$ ions in {the }intergallery spaces. In an aqueous medium, the Na$^+$ ions diffuse out to the bulk {water medium} due to the build-up of an osmotic pressure difference between the intergallery spaces and the bulk. This induces an electrostatic repulsive force between the negatively charged faces of {the clay nano}platelets, {which results in} tactoid swelling and, eventually, {nanoplatelet }exfoliation~\cite{claybook}.
If {the} medium pH$<$11, {the }clay nanoplatelets have weak positively charged rims due to {the} protonation of magnesia groups {and} experience rim-to-face electrostatic attraction, and face-to-face {and} rim-to-rim electrostatic repulsion~\cite{doi:10.1021/acs.langmuir.8b01830,doi:10.1021/acs.langmuir.5b00291}. They also interact {via non-electrostatic} dispersion forces such as van der Waals attraction between their faces and rims~\cite{doi:10.1021/acs.langmuir.8b01830,doi:10.1021/acs.langmuir.5b00291}. 

{Aqueous }Laponite clay suspensions phase{-separate at }concentration{s} below 1\% w/v~\cite{Ruzicka2011} and {self-assemble into }attractive gel{s} with system spanning microstructures {at} clay concentration{s} between 1\% w/v to {4.0\% w/v~\cite{C0SM00590H}}. {Due to tactoid swelling, {nanoplatelet} exfoliation and {the} gradual self-assembly of the heterogeneously charged Laponite {nano}platelets in overlapping coins and house of cards configurations to form system-spanning {gel} networks,  clay suspensions above 1\% w/v undergo physical aging. This leads to a continuous increase in the elastic and viscous moduli of the suspension {with} time as physical aging progresses}~\cite{doi:10.1021/acs.langmuir.8b01830,C0SM00590H, C2SM25731A,D2SM01457B,PhysRevLett.93.228302}. {Increase in} salinity of the aqueous medium accelerates the formation of percolated gel-like networks in the suspension medium~\cite{D1SM00987G,doi:10.1021/acs.langmuir.5b00291,doi:10.1021/acs.langmuir.8b01830}. {A}dding peptizing agents such as tetrasodium pyrophosphate (TSPP) inhibits network formation {as rim-to-face electrostatic attractions reduce considerably due to adsorption of} negatively charged pyrophosphate ions onto the positively charged rims of the Laponite platelet{s}~\cite{MONGONDRY2004191,PhysRevE.66.021401}.  {It has been reported that Laponite suspensions exhibit nematic order {at concentrations above 4\% w/v}~\cite{doi:10.1021/la980117p}.}

This paper proposes a new technique to study the adsorption of charged Laponite nanoplatelets on the surfaces of similarly charged Latex microspheres. Previous experiments had demonstrated that {an optically trapped microsphere executes oscillatory motion due to electrophoresis in an applied {alternating} electric field~\cite{GARBOW2001227,10.1063/1.2884147}. An externally imposed oscillatory electric field of known strength could {therefore}, in principle, be used to measure the electrical force (= effective charge $\times$ electric field) acting on an optically trapped microsphere to extract the effective charge on the {latter}. This technique, {known} as optical tweezer-based single-colloid electrophoresis{,} 
 has been used to determine the charges on colloidal particle{s} and biological samples in polar and non-polar media~\cite{PESCE2014568,10.1063/1.4967401,10.1063/1.2734968,PhysRevLett.108.016101,doi:10.1021/acsnano.3c08161,Grollman:17,10.1063/1.4922039,OKADA200817,GEONZON2023846} {with a} resolution {of a} single elementary charge~\cite{doi:10.1021/acsnano.3c08161,PhysRevLett.108.016101}{,} for measuring the concentration of a target protein in a buffer~\cite{10.1063/1.4922039,OKADA200817} {and} to investigate the adsorption of polymers on a colloidal silica sphere~\cite{C005217P,GEONZON2023846}.

In a previous contribution, some of us {had reported} the adsorption of clay nanoplatelet{s} on {a} trapped Latex microsphere {in} oscillatory active microrheology measurements~\cite{D2SM01457B}. {We had also demonstrated that} the mechanical response of the {microscopically heterogeneous }clay suspension {depends} on {the microsphere diameter}. In the present {work}, we visualise{d} Latex microspheres suspended in {Laponite} clay suspension{s} using cryogenic field emission scanning electron microscopy (cryo-{FE}SEM) {and identified the bright spots on the microsphere surfaces} as adsorbed clay nanoplatelets. The standard deviation of the intensity distribution on the microsphere surface was calculated to quantify the extent of nanoplatelet adsorption. The addition of sodium chloride (NaCl) or TSPP in the suspension {medium} was seen to significantly influence the adsorption {process. T}he kinetics of Laponite nanoplatelet adsorption on {a} trapped Latex microsphere in {an }aqueous medium {were next estimated} using optical tweezer-based single-colloid electrophoresis experiments. Adsorption of clay nanoplatelets results in the transfer of net charges to the microsphere surface. We characterise the nanoplatelet adsorption process by measuring the temporal evolutions of the effective surface charges {and the hydrodynamic drags }on microspheres trapped in clay suspensions with different salt contents. 
We observed {that} the effective surface charge {and hydrodynamic drag show similar increasing trends} with increasing clay concentration, NaCl concentration and time. We identified two distinct mechanisms {that drive the adsorption process. Besides} rapid adsorption due to non-electrostatic dispersion forces, {we also observe{d} slower and more gradual} adsorption, driven by the weakening of electrostatic repulsion between the microsphere and the faces of the {self-assembled }Laponite nanoplatelets at later times. 

{We} successfully {tuned} the adsorption {of Laponite clay on a Latex microsphere} by incorporating controlled amounts of NaCl or TSPP in the suspension medium. The extent of nanoplatelet adsorption, as quantified from cryo-FESEM images, agrees with {our }effective surface charge measurement{s extracted} from optical tweezer-based single-colloid electrophoresis experiments. The findings of our study are useful in stabilising colloidal clay-armoured Latex particles in various scenarios, such as during Pickering emulsion polymerization~\cite{doi:10.1021/acs.langmuir.6b01080,doi:10.1021/acs.langmuir.5b03576} {and in estimating} the real-time adsorption efficiency of patchy particles~\cite{C0CP02296A}.

\section{\label{em}Materials and methods}
	\subsection{Sample preparation}
Experiments were performed with aqueous suspensions of Laponite$^\text{\textregistered}$ XLG (BYK Additives Inc.), a synthetic clay powder constituted by disk-shaped clay nanoplatelets with a diameter of $25$-$30$ nm and thickness of $0.92$ nm. Since Laponite$^\text{\textregistered}$ is hygroscopic, the absorbed moisture was removed by baking the clay powder in an oven for $18$-$24$ hours at  $120^{\circ}$C. Slighly negatively charged polystyrene Latex microspheres were procured from Bangs Laboratories, Inc. NaCl (LABORT Fine Chem Pvt. Ltd) or TSPP (E. Merck (India) Ltd) were mixed in {ultrapure} Milli-Q water (Millipore Corp., resistivity = $18.2$ M$\Omega$-cm) to prepare aqueous solutions of {pre-determined} concentrations. 
A fixed quantity of dried Laponite$^\text{\textregistered}$ powder was added to 50 ml {MilliQ} water or to {NaCl or TSPP} solutions prepared as above. The mixture{s were} stirred continuously for 40 minutes to prepare homogeneous suspension{s}. 10 ml of {a} freshly prepared suspension {was} filtered using a syringe filter of pore size 0.45 $\mu$m (SLHAR33SS Millex). A very dilute {suspension} of Latex microspheres of diameter 1.0 $\mu${m} was added to the filtered Laponite$^\text{\textregistered}$ suspension and stirred for another 5 minutes. The suspensions {thus prepared} were used for {cryo-FESEM} and optical tweezer electrophoresis measurements.

\subsection{Cryogenic field emission scanning electron microscopy (cryo-FESEM)}
Freshly prepared Laponite$^\text{\textregistered}$ suspensions with added Latex {microspheres} were loaded in capillary tubes (Capillary Tube Supplies Ltd, UK) of diameter 1 mm. The tubes were subsequently sealed at their two ends and 
left undisturbed for 90 minutes at room temperature (24$\pm$2$^{\circ}$C). The instant at which the tube {was} sealed was recorded as the start of the experiment (aging time $t_{w}$ = 0). The samples {in the tubes} were {next} vitrified in liquid nitrogen slush at -190$^{\circ}$C (PP3000T Quorum technology) and transferred to a vacuum chamber. A precision knife was used to cut the sample, which was then sublimated for 15 min at -90$^{\circ}$C. {A g}old coating of thickness $\approx$ 5 nm was applied to the sample surface to ensure good image contrast. Finally, the samples were transferred to the cryo-chamber, maintained at a temperature of -190$^{\circ}$C, for imaging. Back-scattered secondary electrons were used to reconstruct the surface images of the samples. {Raw grayscale cryo-FESEM images of trapped Latex micropheres are displayed} in Fig.~S1 of the supplementary information. {Python, version 3.9.13, in Jupyter Notebook was used to characterise t}he extent of adsorption of clay nanoplatelets on the microsphere by calculating the standard deviation {of} the {pixel} intensity distribution of the {cropped} grayscale microsphere image. A statistical analysis of the intensity profile {of an image of a microsphere embedded in a 2.5\% w/v Laponite suspension at $t_{w}$ = 90 mins} (Fig.~S2) reveals that a majority of the bright spots in the grayscale cryo-FESEM image, including the ones seen on the microsphere surface, correspond to individual Laponite clay{ nanoplatelets and small aggregates}.  

 \begin{figure}[t]
     \centering
     \includegraphics[width=1\linewidth]{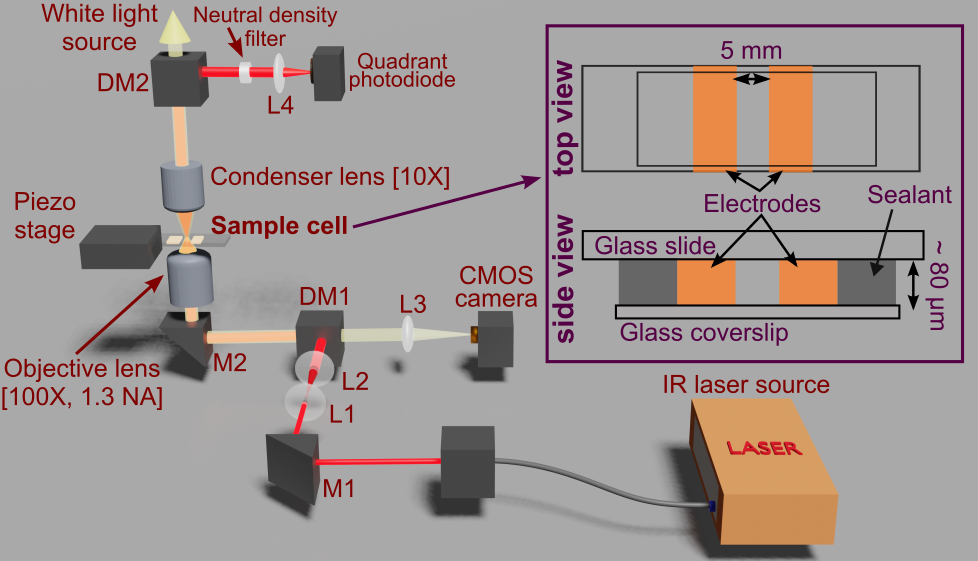}
     \caption{Schematic illustration of the optical tweezer setup used for single-colloid electrophoresis experiments. M1 and M2 are mirrors, DM1 and DM2 are dichroic mirrors, L1, L2, L3 and L4 are {achromatic doublet} lenses of focal lengths 35 mm, 150 mm, 200 mm and 35 mm respectively. The infrared laser beam is indicated in red. (Inset) Schematic illustrations of the top and side views of the sample cell.}
     \label{fig:1}
 \end{figure}
\subsection{Sample cell for optical tweezer-based single-colloid electrophoresis}
The sample cell, schematically illustrated in the inset of Fig.~\ref{fig:1}, was prepared using a glass slide, a \#1.0 coverslip (thickness $\approx$ 150 $\mu$m, Blue Star, India), and two copper strips, each of thickness $\approx$ 80 $\mu$m. The copper strips act both as electrodes and spacers. The glass slides and coverslips were adhered to each other using {a m}ulti-purpose sealant (Dowsil 732). Since each sample cell {had} a slightly different electrode separation, $d$, we measured the distance between {the} electrodes in each cell using a digital vernier calliper ($\Delta d = $ 0.01 mm) and reported  $4.80$ mm $<d$ $<5.30$ mm for {the} different sample cells {used in this work}. 
\subsection{Optical tweezer setup}
Figure~\ref{fig:1} schematically illustrates the optical tweezer (OT, ThorLabs OTKB module) setup used to perform single-colloid electrophoresis experiments. A linearly polarized infrared laser (YLR-5-1064-LP, IPG Photonics) with a wavelength of $1064$ nm and a Gaussian beam profile (TEM00 mode) was used for trapping. After reflecting from the guiding mirror (M1), the beam was expanded using a combination of two lenses (L1 and L2) to overfill the back aperture of the objective lens. The expanded beam passed through the dichroic mirror (DM1) and was guided to the oil immersion objective (100X, 1.3 NA Olympus) to form a diffraction-limited spot. The gradient forces produced by the focussed laser beam trapped the microsphere in three dimensions. The scattered beam was collected using a condenser lens (10X, Nikon). The reflected beam from the dichroic mirror (DM2) was attenuated by a neutral density filter and focussed using the lens (L4) on a quadrant photodiode (QPD - PDQ80A, Thorlabs). The quadrant photodiode (QPD) was kept at the back focal plane of the condenser and used to detect nanometer-scale deflections of the trapped Latex microsphere~\cite{10.1063/1.2356852,10.1063/1.1785844}. The output from the QPD was recorded using Labview 2021 with a data acquisition device (DAQ, NI USB-6218, National Instruments) at a sampling rate of 30 kHz. The protocols 
{for QPD calibration and conversion of} microsphere deflection{s} to power spectral density (PSD) of the position fluctuations of the trapped microsphere were implemented using MATLAB 2024a as described in Figs.~S3 and S4 of the supplementary information. The stiffnesses of the optical trap at various laser powers, determined by trapping Latex microspheres in water in the absence of an electric field, are displayed in Fig.~S5 of the supplementary information. In this {work}, the laser power used was $330$ mW and the measured trap stiffness was $50.61$ {pN/$\mu$m} $\pm$ $0.37$ pN/$\mu$m.

 For each single-colloid electrophoresis experiment, a freshly prepared Laponite clay suspension with Latex microspheres was loaded in the sample cell which was {then} placed on a piezo-controlled stage. The electrodes of the sample cell were connected to a frequency generator (Agilent 33220A). A sinusoidal electric field was applied at $t_w =$ 0 minutes and maintained throughout the experiment. As the viscoelasticity of the clay suspension increased rapidly with clay concentration and age, the position fluctuations of the trapped microsphere became increasingly more difficult to resolve, rendering the trap ineffective. We chose the concentration and age ranges of the Laponite suspension media accordingly, and retained only those experiments wherein the corner frequency, $f_c$, of the PSD of microsphere position fluctuations was greater than 100 Hz.
 

 \subsection{Measurement of {the} effective surface charge on the Latex microsphere}
The motion of a microsphere in an optical trap in the presence of a uniform AC electric field, $E(t) = E_0 \sin(2\pi f_{AC}t)$ applied along the $\hat{x}$ direction (Fig.~\ref{fig:2}(a)), is described by the Langevin equation for a periodically driven Brownian oscillator~\cite{PESCE2014568,10.1063/1.4922039,B703994H}:
\begin{equation}
    m\Ddot{x}(t) + \gamma \left[\Dot{x}(t) - v_{eo}\right] + \kappa x(t) = F_T(t) + F_E(t)
\end{equation}
 Here, $m$ is the mass of the trapped microsphere, $x(t)$ is the $\hat{x}$ component of the centre of mass motion of the microsphere, $\gamma$ is the {Stokes} drag coefficient, $v_{eo}$ is the electro-osmotic flow velocity of the medium, $\kappa$ is the trap stiffness, {$F_E$ is the electrical force} and $F_T(t) = \sqrt{2\gamma k_BT}\xi(t)$ is the random thermal force. In the expression for $F_T(t)$, $k_B$ is the Boltzmann constant, $T$ is the absolute temperature and $\xi(t)$ is the zero-mean delta-correlated white noise with $<\xi(t)> = 0$ and $<\xi(t)\xi(t^\prime)> = \delta(t - t^{\prime})$. As discussed earlier, we {trapped the microsphere in} clay suspensions {of} small ages to ensure trap effectiveness. Since these samples were predominantly liquid-like, the zero-mean delta-correlated white noise approximation holds for our experiments. In addition, as the appl{ied} electric field was expected to rupture some of the fragile microstructures in the clay suspension, the {optically trapped} Latex microsphere was modelled as a Brownian particle.
 
 For an applied AC {electric} field of strength $E_0$ and frequency $f_{AC}$, $F_E(t) = Q_{eff}E_0\sin({2\pi f_{AC} t + \phi})$, where $Q_{eff} = eZ_{eff}$ is the effective charge on the {surface of the} trapped microsphere and $\phi$ is the phase difference between the centre of mass position of the microsphere and the applied field. In the expression for $Q_{eff}$, $e$ is the elementary charge and $Z_{eff}$ is the effective number of elementary charges on the trapped microsphere.
  A previous study has shown that at sufficiently high AC frequencies, the contribution of electro-osmosis to the motion of a trapped Latex microsphere {can} be neglected~\cite{SEMENOV2009260}. 
 We have applied an AC electric field of frequency $f_{AC} = 7.992$ kHz while trapping the microsphere at least 7 $\mu$m away from the surface. 
 Considering $v_{eo} \approx 0$ and an overdamped limit, $m\Ddot{x}(t) = 0$ since Reynold's number $<10^{-4}$ in all our experiments, a Fourier transform of Eqn.~(1) yields the one sided power spectral density, $S_{xx}(f)$, which describes the oscillation amplitude of the trapped microsphere:~\cite{PESCE2014568,10.1063/1.4922039,B703994H}:
\begin{equation}
    S_{xx}(f) = \frac{k_BT}{\pi^2 \gamma} \frac{1}{f^2 + {f_c}^2} + \frac{k_BT}{\kappa} \Gamma^2 \delta(f - f_{AC}) 
\end{equation}
The first term in Eqn.~(2) is a Lorentzian function with a corner frequency $f_c = \kappa/{2\pi\gamma}$ and represents the Brownian dynamics of the trapped microsphere. The second term in Eqn.~(2) represents the oscillatory motion of the microsphere in the alternating electric field and is characterised by a $\delta$-peak at $f = f_{AC}$. The dimensionless parameter $\Gamma^2$ is proportional to the ratio of the mean-square electrical and Brownian forces:~\cite{PESCE2014568,10.1063/1.4922039,B703994H}
\begin{equation}
    \Gamma^2 = \frac{1}{1 + \left(f_{AC}/f_c\right)^2}\frac{e^2{Z_{eff}}^2{E_0}^2}{2k_BT\kappa}
\end{equation}

Equation~(3) can be rearranged to yield the effective surface charge on the trapped microsphere~\cite{B703994H}: 
\begin{equation}
    eZ_{eff} = \frac{\Gamma}{E_0}\sqrt{2 k_B T \kappa\left[ 1 + \left(\frac{f_{AC}}{f_c}\right)^2 \right]}
\end{equation}
\begin{figure*}[ht]
     \centering
     \includegraphics[width=1\linewidth]{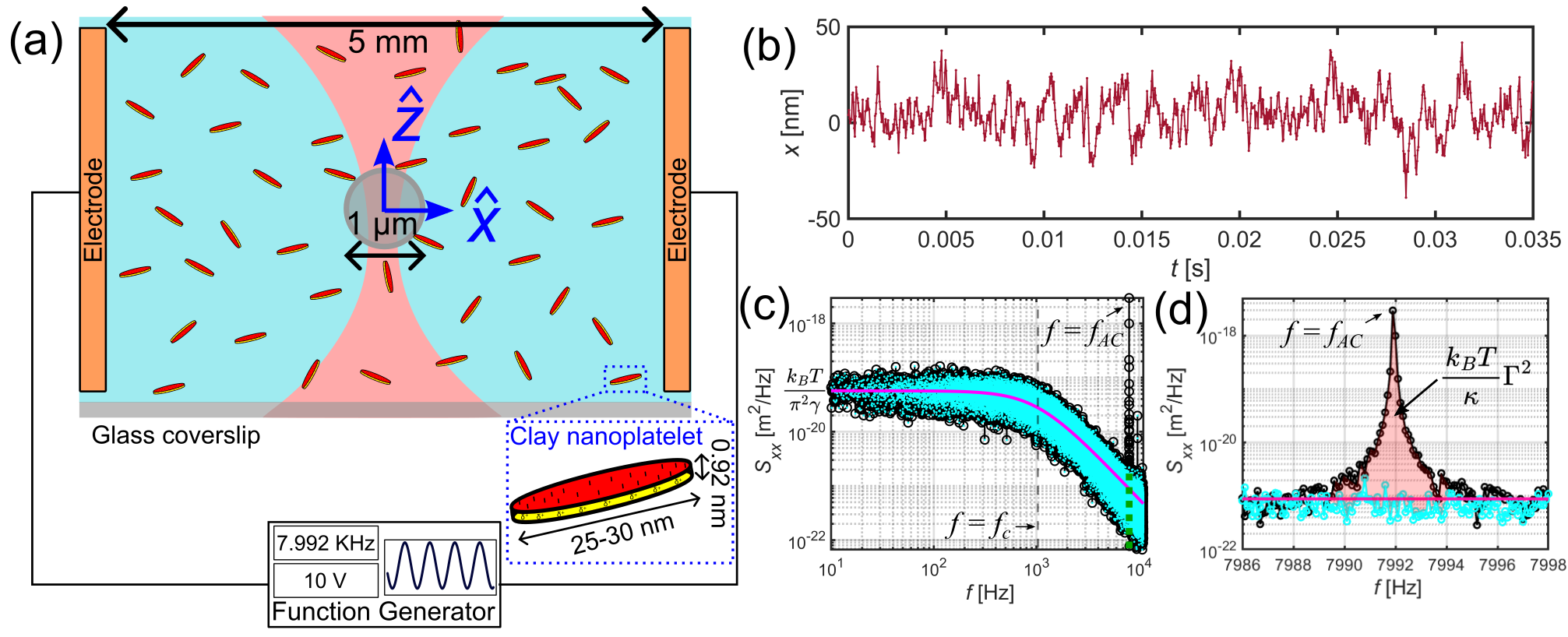}
     \caption{Schematic illustration of {a} slightly negatively charged Latex microsphere of radius 1.0 $\mu$m {optically trapped} in a {Laponite} clay suspension. The {AC} electric field was applied along the $\hat{x}$ direction. (b) A representative trajectory of {a} microsphere trapped in water {in} an {AC} electric field. (c) Power spectral density, $S_{xx}(f)$, and (d) {a} zoomed-in {view} of the peak {centred} at $f_{AC} = 7.992$ kHz. The dimensionless parameter $\Gamma^2$ is computed from the area under the peak {shown in pink}.} 
     \label{fig:2}
 \end{figure*}

We have computed the Fourier transform of the trajectory of the trapped microsphere, $x(t)$, obtained from our single{-}colloid electrophoresis experiment (representative data {displayed in  Fig.~\ref{fig:2}(b)}), to obtain the PSD, $S_{xx}(f)$ (Fig.~\ref{fig:2}(c)). Drag-dominated Brownian forces overcome optical forces at frequencies higher than $f_c$, {therefore,} $S_{xx}(f) \sim 1/f^2$ at $f>f_c$. Furthermore, an {additional} contribution to the PSD, arising from microsphere oscillation in the applied AC electric field, is clearly visible as a sharp peak at the {applied} AC frequency $f_{AC} = 7.992$ kHz (green dashed line in Fig.~\ref{fig:2}(c)). Fig.~\ref{fig:2}(d) shows a zoomed-in view of the peak centred around $f_{AC}=7.992$ kHz. {Electric power is {therefore} distributed {with}in a narrow frequency range {under the peak, the finite height and width of which} (pink shaded area in Fig.~\ref{fig:2}(d)) {arises} from the discrete data acquisition process}. Following previous work, we computed the electric power by estimating the area under the peak{~\cite{10.1002/elps.201300214}} at $f=f_{AC}$ (pink shaded area in Fig.~\ref{fig:2}(d)) using Eqn.(2). 

The corner frequency, $f_c$, was estimated by fitting a Lorentzian function to $S_{xx}(f)$ {(Fig.~\ref{fig:2}(c))}. {The areas under the sharp peaks were} calculated using the trapezoidal numerical integration procedure, after subtracting the thermal baseline (red line in Fig.~\ref{fig:2}(d)). {Since trap stiffness $\kappa$ and room temperature $T$ are known, the electrical power transferred during the adsorption process was calculated using Eqn.~(2).} In all our electrophoresis experiments, the oscillatory motion of the trapped microsphere {in the applied electric field} was {substantially} smaller than its overdamped Brownian motion in the viscous medium, therefore $\Gamma{^2} < 1$ (Fig.~\ref{fig:3x}(a)). 
Finally, the {Stokes} drag coefficient, $\gamma = \kappa/2 \pi f_c$, was also calculated.
 \begin{figure}[t]
     \centering
     \includegraphics[width=1.0\linewidth]{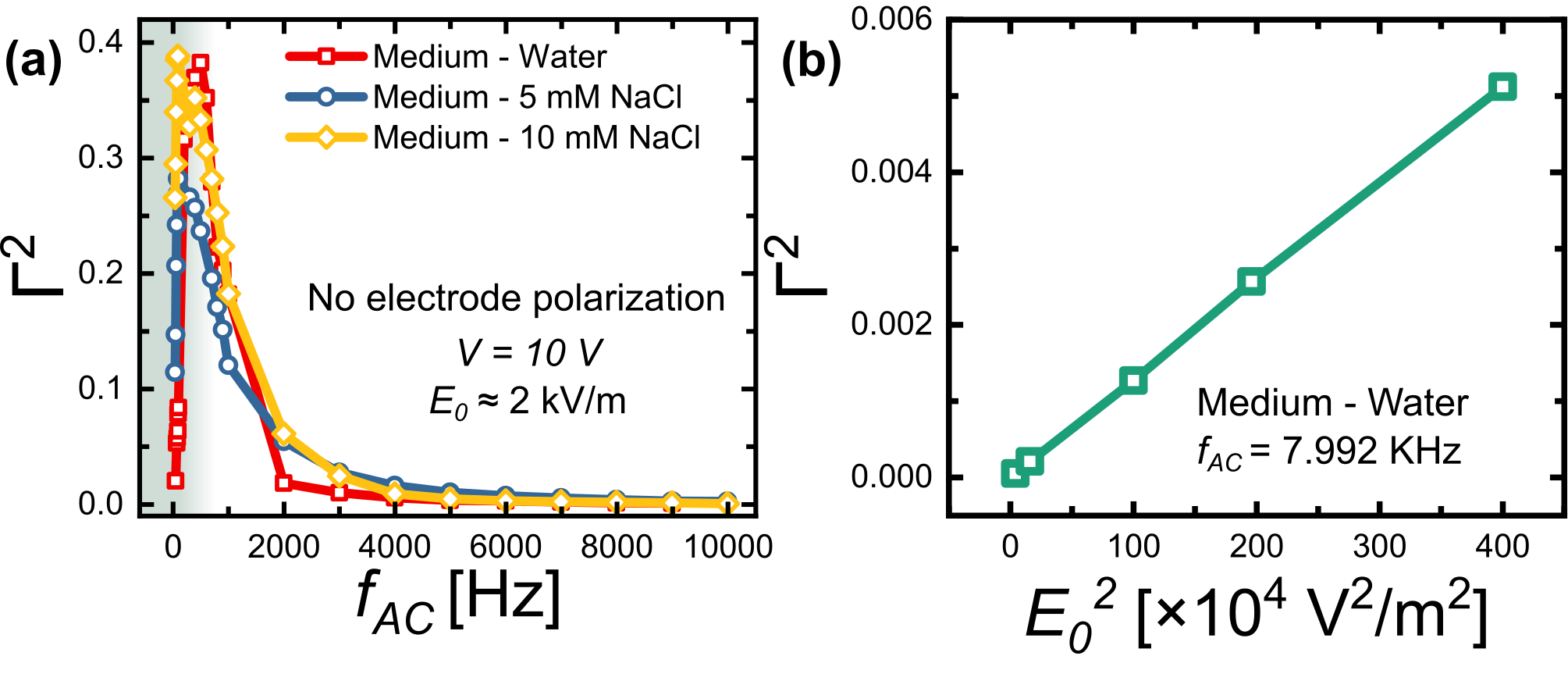}
     \caption{(a) Dimensionless parameter $\Gamma^2$, proportional to the ratio of mean-square electric and Brownian forces, as a function of AC field frequency, $f_{AC}$, for a microsphere trapped in aqueous media {with} different ionic {contents}. (b) $\Gamma^2$ as a function of the square of electric field strength, ${E_0}^2$, for the microsphere trapped in water.}
     \label{fig:3x}
 \end{figure}

Before {acquiring data for} single-colloid electrophoresis of a microsphere trapped in clay suspensions, we validated our setup by trapping the microsphere in {an} aqueous medi{um}. We measured the dimensionless parameter $\Gamma^2$ as a function of electric field {strength}, $E_0$, and frequency, $f_{AC}$. The observed frequency-dependent variation of $\Gamma^2$ at a fixed {AC} electric field is plotted in Fig.~\ref{fig:3x}(a) for different ionic contents {in an aqueous medium}. The initial increase in $\Gamma^2$ at low $f_{AC}$ arises from a reduction in the electrode polarisation-induced screening of the applied electric field~\cite{10.1002/elps.201300214}. 
The trapped microsphere experienced enhanced viscous damping when $f_{AC}$ was increased, with $\Gamma^{2}$ decreasing to values $\sim$ O($10^{-3}$) when $f_{AC} \sim$ O(kHz). In Fig.~\ref{fig:3x}(b), $\Gamma^2$ of {a} microsphere trapped in water is plotted {\textit{vs.} {$E_{\circ}^2$} at $f_{AC} = 7.992$ kHz. As {predicted by Eqn.~(3)}, {a} linear dependency of $\Gamma^2$ on $E_{0}^{2}$ is observed. For {our {subsequent} electrophoresis} experiments, we chose $E_0 \approx 2$ kV/m.

	\section{\label{r&d}Results and Discussion}
	
	\begin{figure}[ht]
     \centering
     \includegraphics[width=0.75\linewidth]{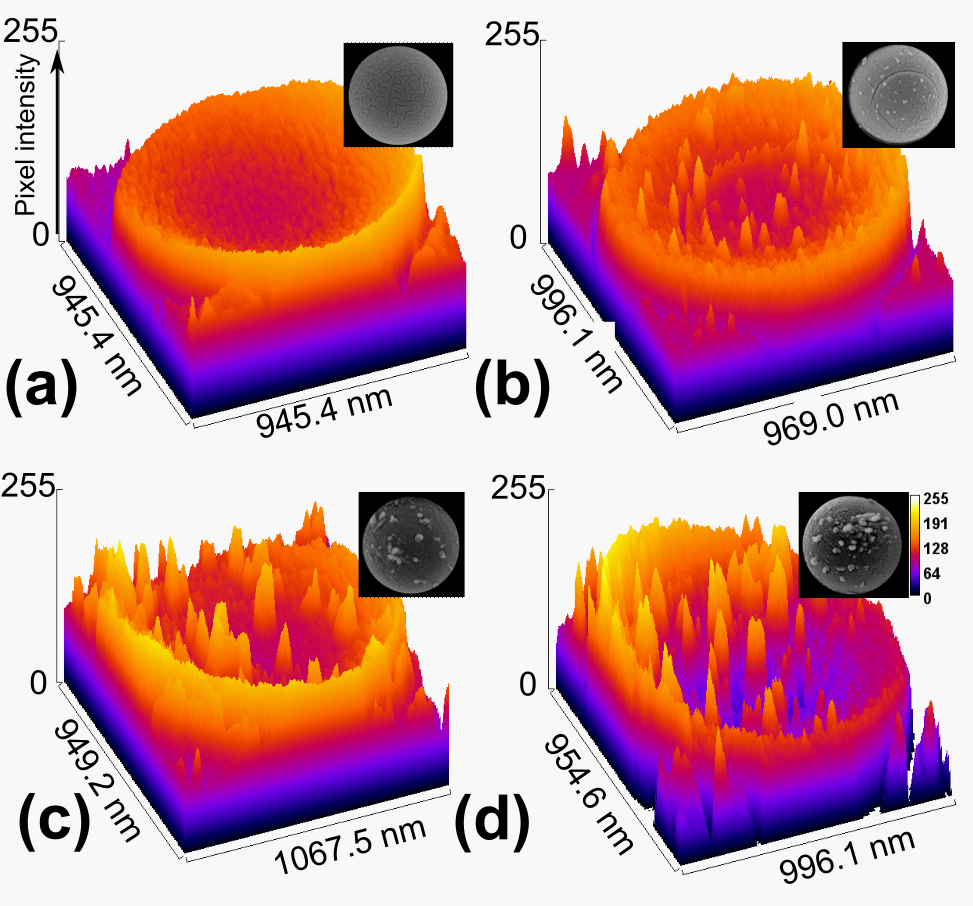}
     \caption{Cropped cryo-FESEM images {(insets)} and the corresponding pixel intensity variation maps of the surfaces of Latex microspheres of diameter 1.0 $\mu$m in (a) water, (b) 2.5\% w/v aqueous Laponite suspension, (c) 2.5\% w/v aqueous Laponite suspension in the presence of 1.25 mM TSPP and (d) 2.5\% w/v aqueous Laponite suspension in the presence of 1.25 mM NaCl at an aging time $t_w = $ 90 minutes. 
     }
     \label{fig:3}
 \end{figure}
\subsection{Observation of clay nanoplatelet adsorption on a Latex microsphere using Cryo-FESEM}
{The physical aging behaviour of Laponite clay suspensions has been quantified {extensively} in {several} rheological studies by measuring the time-}evolution of suspension viscoelasticity both in the presence and absence of externally added salts~\cite{doi:10.1021/acs.langmuir.5b00291,D1SM00987G}. It was seen that while clay aging is fastest in the presence of NaCl, it {could} be completely suppressed {by adding an} adequate quantity of TSPP {to the aqueous medium}~\cite{PhysRevE.66.021401,2024arXiv240701396S}.

 {To verify the adsorption of {Laponite clay nanoplatelets on Latex microsphere surfaces} in different ionic conditions, we first conducted direct visualisation {via} cryo-FESEM.} All the cryo-FESEM experiments displayed here were performed at a fixed aging time, $t_w =$ 90 minutes. Grayscale cryo-FESEM images of microspheres embedded in water or clay suspensions {in} various {ionic} conditions are displayed in Fig.~S1. We cropped these images using a masking algorithm in Python to focus on the adsorption of clay nanoplatelets on the surfaces of the Latex microspheres. Pixel intensity variations obtained from the grayscale images of the microsphere surface, with the corresponding cropped images in the inset, are presented in Fig.~\ref{fig:3}. When the microsphere was suspended in water, a smooth surface texture was observed, as displayed in Fig.~\ref{fig:3}(a). When the microsphere was embedded in a clay suspension of concentration 2.5\% w/v, several bright spots were observed on its surface, as displayed in Fig.~\ref{fig:3}(b). The histogram of the size distribution of the bright spots is plotted in Fig.~S2(b) of the supplementary information.  The histogram shows peaks at $\approx$ 25 nm {and 50 nm}, which correspond {respectively} to the lateral dimensions of individual Laponite nanoplatelet{s and small particle clusters.} 

  \begin{figure}[th]
     \centering
     \includegraphics[width=0.65\linewidth]{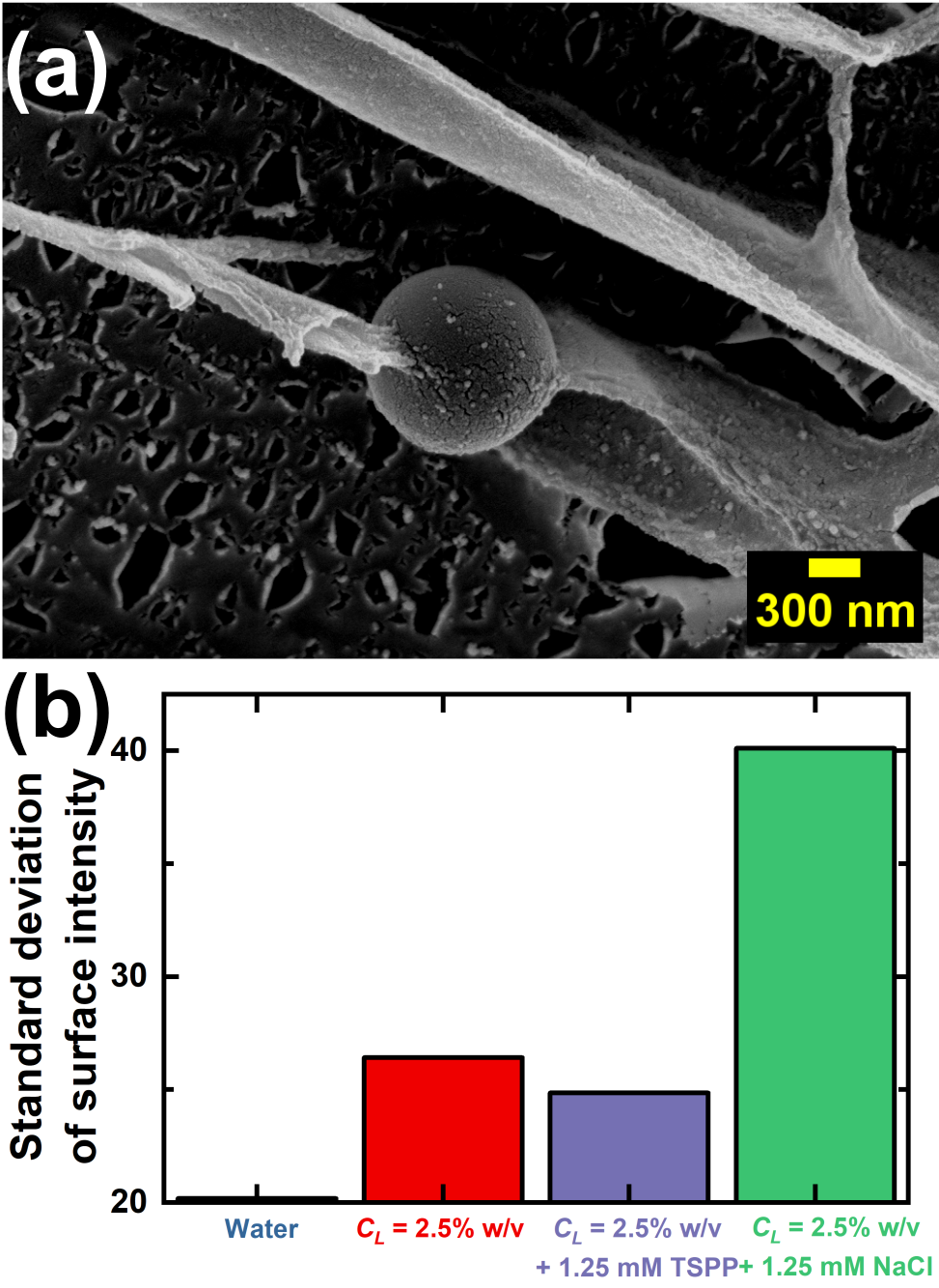}
     \caption{(a) Cryo-FESEM image of a Latex microsphere attached to the strands of an evolving {Laponite} clay gel network. (b) Standard deviations of surface intensities {in arbitrary units for} Latex microspheres embedded in water or clay suspensions.
     }
     \label{fig:3-2}
 \end{figure}
  
 Fig.~\ref{fig:3}(c) shows the intensity variation map on the microsphere surface due to the adsorption of clay nanoplatelets in the presence of tetrasodium pyrophosphate (TSPP). It is known that the attachment of pyrophosphate ions ({P$_2$O$_7$}$^{4-}$) to the rims of {Laponite clay} nanoplatelets {can reduce or neutralise the positive charges or even render negative charges on the rims~\cite{PhysRevE.66.021401}. 
 {E}nhanced adsorption of clay nanoplatelets {is} clearly evident {from Fig.~\ref{fig:3}(c) for} this system, which suggests that non-electrostatic dispersion forces initiate the adsorption process. In the presence of NaCl, an increase in the number of adsorbed nanoplatelets on the microsphere surface is evident from Fig.~\ref{fig:3}(d). 
 A higher number density of Na$^+$ and Cl$^-$ ions in the suspension medium screens the electrostatic repulsion between the nanoplatelets and microsphere. As a consequence, there is a relatively higher probability of a nanoplatelet approaching very close to (within O($10$ nm) of) the microsphere surface. Under these conditions, {acceleration in the} Laponite adsorption {process} is expected due to short-range non-electrostatic dispersion interactions.  It was shown previously that increasing NaCl concentration in the medium leads to an increased rate of adsorption of nanoplatelets on a {Latex} microsphere {and the eventual} formation of {nanoplatelet} multilayers~\cite{doi:10.1021/acs.langmuir.6b01080,doi:10.1021/acs.langmuir.5b03576}. 
 As discussed earlier, {Laponite} clay suspensions form a gel at concentrations above 1\% w/v~\cite{doi:10.1021/acs.langmuir.8b01830}. Interestingly, we observed in an earlier study that the evolving gel {networks} can also attach to the trapped microsphere \cite{D2SM01457B}. A representative image is shown in Fig.~\ref{fig:3-2}(a).

       
{In summary,} we determined that {the} {dominant} mechanism driving the adsorption process is non-electrostatic dispersion forces. Since the adsorbed nanoplatelets appear as bright spots, the standard deviation in intensity variation profiles on the surface of the microsphere, as displayed in Fig.~\ref{fig:3}, should give us {valuable information on} the extent of nanoplatelet adsorption. We {therefore quantified} the adsorption of Laponite nanoplatelets in different {ionic} conditions of the suspension medium by computing the standard deviations of the {pixel} intensities on the surfaces of the Latex microspheres.  {Representative bar plots are displayed in }Fig.~\ref{fig:3-2}(b). A higher standard deviation of the pixel intensities in the presence of NaCl confirms a very high degree of nanoplatelet adsorption. 

\subsection{Quantifying the kinetics of nanoplatelet adsorption on a trapped Latex microsphere using single-colloid electrophoresis measurements}
\begin{figure*}[ht]
     \centering
     \includegraphics[width=1\linewidth]{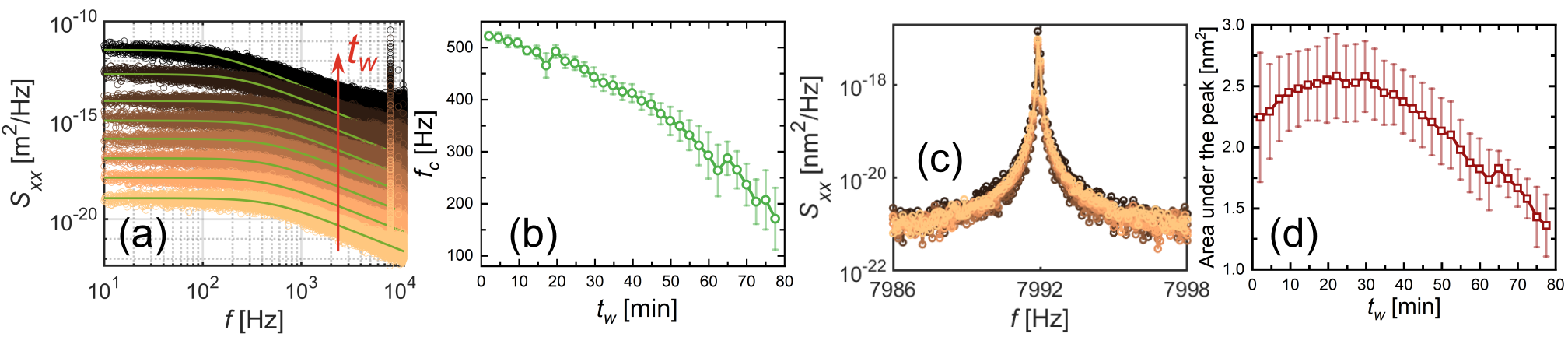}
     \caption{(a) Power spectral densities (PSDs), $S_{xx}(f)$, of a Latex microsphere optically trapped in a Laponite clay suspension of concentration 2.5\% w/v for various aging times, $t_w$.  The red arrow points towards increasing aging times. $S_{xx}(f)$ for $t_w$ = 12, 22, 32, 42, 52, 62 and 72 minutes were scaled for better visualisation.  (b) The corner frequencies, $f_c$, calculated from Lorentzian fits to the PSDs, {\textit{vs.}} $t_w$. (c) The zoomed-in peaks in $S_{xx}(f)$ at $f=f_{AC}$ for various aging times, $t_w$. The colour codes for the samples of different $t_w$ are the same as in (a). (d) Area{s} under the peak {in $S_{xx}(f)$} centred at $f_{AC}$ = 7.992 kHz {\textit{vs.}} $t_w$. Error bars in (b,d) are standard errors from four independent measurements.}
     \label{fig:4}
 \end{figure*}
We next {quantified} the clay adsorption process by implementing optical tweezer electrophoresis.  
Since the adsorption of {Laponite clay nanoplatelets} {transfers} net charges {to the surface} of a Latex microsphere, we measured the surface charge of a single {optically trapped} microsphere as a function of {clay} suspension aging time, $t_w$, to gain insights into the kinetics of the adsorption process. Thermal fluctuations of optically trapped microsphere{s} in Laponite clay suspension{s} were {monitored} to simultaneously measure the effective surface charge {and Stokes drag on} the former. Figure~\ref{fig:4}(a) shows the power spectral densities, $S_{xx}(f)$, calculated from the experimental trajectories of a microsphere {trapped} in an aqueous clay suspension of concentration 2.5\% w/v, {vs.} aging time, $t_w$. The PSD curves at $t_w = $ 12, 22, 32, 42, 52, 62 and 72 minutes were shifted vertically for better visuali{s}ation. $S_{xx}(f)$ exhibits the Lorentzian functional form as a function of frequency, besides also exhibiting a prominent peak at $f = f_{AC}$ for all {$t_{w}$}. {We see from Fig.~\ref{fig:4}(b) that} the corner frequency,  $f_c = \kappa/2 \pi \gamma$,  of the optical trap decrease{s} with increasing aging time, $t_w$, of the clay suspension. Since the change in the refractive index of the clay suspension during physical aging is expected to be negligible~\cite{RAVIKUMAR2008326}, the decreasing value of corner frequency indicates increased drag, $\gamma$, on the trapped microsphere due to the {evolution} of {gel networks} in the aging clay suspension. 

 We {next} focus on the peaks in $S_{xx}(f)$ {centred} at $f_{AC} = 7.992$ kHz for various clay suspension ages, $t_{w}${,} in Fig.~\ref{fig:4}(c). The electric power transferred to the microsphere is equal to the area under the peak and decreases with increasing clay suspension age, $t_{w}$, at later times, as shown in Fig.~\ref{fig:4}(d). 
 {At} $f > f_{c}$, viscous forces are expected to dominate the electric forces~\cite{PhysRevLett.108.016101}}. The decrease in electric power with increasing $t_w$, observed in Fig.~\ref{fig:4}(d), arises from increase in viscous damping of the trapped microsphere as the suspension viscoelasticity increases due to {the gradual self-assembly of clay gel networks} during the physical aging process.
 \begin{figure*}[ht]
     \centering
     \includegraphics[width=1\linewidth]{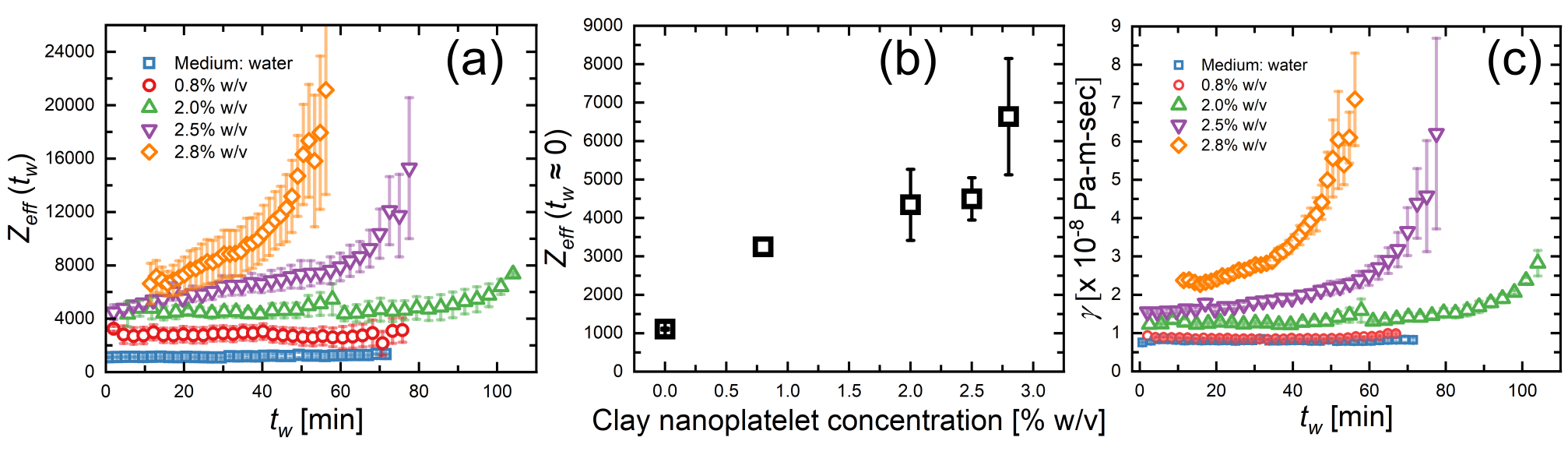}
     \caption{(a) $Z_{eff}(t_w)$ as a function of aging time, $t_w$, for various {Laponite }clay nanoplatelet concentrations, $C_L$. (b) $Z_{eff}${($t_w\approx0$}) as a function of $C_L$ and (c) the {Stokes drag coefficient}, $\gamma(t_w)$, characterising the trapped microsphere motion as a function of $t_w$ for various $C_L$. {Error bars are standard errors from three independent measurements.}}
     \label{fig:5}
 \end{figure*}

We calculated the effective charge, $Q_{eff} = eZ_{eff}$, on the trapped microsphere using Eqns.~(2,4) and our experimentally evaluated values of $f_c$ and $\Gamma$ for various Laponite concentrations, $C_L$, and aging times, $t_w$. 
The {time-}evolution {behaviours of {the effective number of elementary charges, }}$Z_{eff}$, {are plotted in Fig.~\ref{fig:5}(a)} as a function of $t_w$ {for different $C_L$}. No temporal changes in $Z_{eff}$ were seen when the microsphere was trapped in water {and in a dilute clay suspension of $C_L$ = 0.8\% w/v}. {{Interestingly, }when trapped in the Laponite clay sample of $C_L$ =} 0.8\% w/v, the measured $Z_{eff}(t_w \approx 0)$ of the microsphere {was seen to increase} by a factor of three (Fig.~\ref{fig:5}(b)) {when compared to its $Z_{eff}$ measured} in a water medium.
We attribute the observed increase in $Z_{eff}(t_w \approx 0)$ to the transfer of net charges  {from the adsorbed} clay nanoplatelets {to} the surface of the microsphere. {It can be seen} from Fig.~\ref{fig:5}(b) that $Z_{eff}(t_w\approx0)$ increased {further when $C_L$ was increased}. {This observation can be understood by considering accelerated Laponite {nanoplatelet} adsorption due to {their} higher availability {in the suspension medium}}. {Furthermore, it is seen from Fig.~\ref{fig:5}(a) that} the measured $Z_{eff}(t_w)$ increases with increasing aging time, $t_w$, for {the higher $C_L$ values of 2.0, 2.5 and }2.8 \% w/v. We note that the increase in $Z_{eff}(t_w)$ becomes more rapid {with increasing $C_L$}. {At higher $C_L$ and $t_w$, higher rates of Na$^+$ dissolution from the faces of Laponite nanoplatelets lead to a greater reduction in the strength and range of the electrostatic repulsion between like-charged surfaces (clay nanoplatelet faces and microsphere surface) in the aqueous medium~\cite{doi:10.1021/acs.langmuir.5b00291}.}
As a result of enhanced electrostatic screening at higher $C_L$, the formation of clay aggregates 
{and clay} gel networks get significantly accelerated. The adsorption of {single} clay nanoplatelets, small clay aggregates and network strands on the microsphere surface speeds up under these conditions, as manifested by the stronger temporal evolution of $Z_{eff}(t_w)$ with time. 

{A nanoplatelet of Laponite clay has around -700 elementary charges on its faces~\cite{10.1063/1.480582}. The charges on the rims are pH-dependent and reported to be +50 elementary charges at a pH of 9.97 at 25$^\circ$C~\cite{TAWARI200154}.  From Fig.~\ref{fig:5}(a), we record a change in $Z_{eff}$ $\sim$ O$(10^4)$ over the entire experimental duration. Since each Laponite nanoplatelet has O(10$^{2}$) charges on its surface, our best estimate for the effective number of adsorbed nanoplatelets is $O(10^{2})$, which suggests very weak adsorption. 
 We next calculated the {Stokes drag coefficient, $\gamma$, for the trapped Latex microsphere}, as a function of aging time, $t_w$, and plotted the data in Fig.~\ref{fig:5}(c). 
 We see from Eqn.~(4) that for {$f_{AC}>f_{c}$}, $Z_{eff}$ and $\gamma$ are both proportional to ${1/f_c}$. As expected under these conditions, the time-evolution of $\gamma$, {as} seen in Fig.~\ref{fig:5}(c), shows approximately the same temporal variation as noted earlier for $Z_{eff}$ in Fig.~\ref{fig:5}(a).}

\begin{figure*}[ht]
     \centering
     \includegraphics[width=1\linewidth]{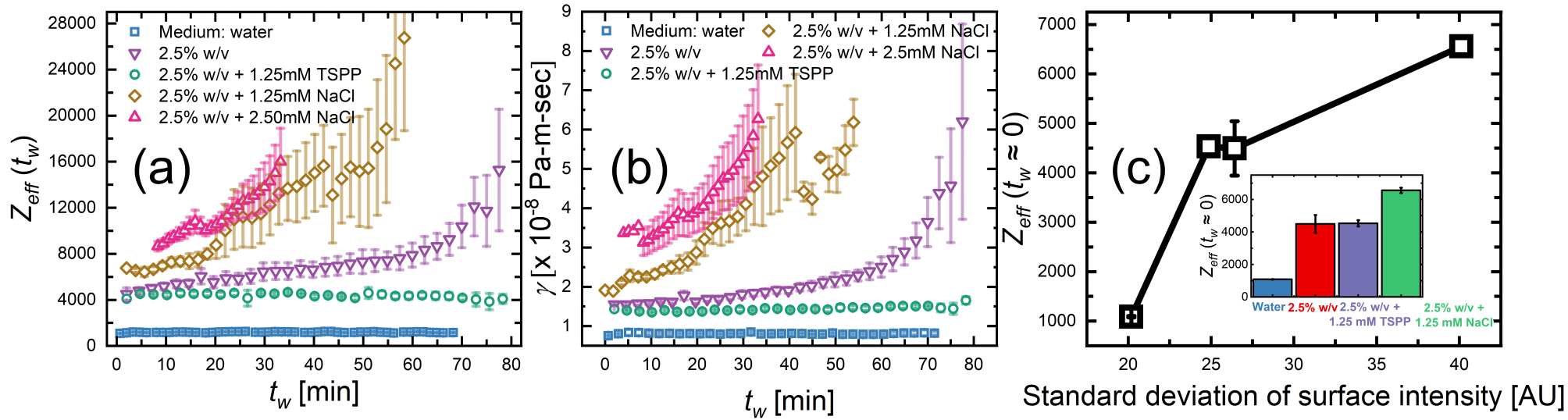}
     \caption{(a) $Z_{eff}(t_w)$ and (b) Stokes drag coefficient $\gamma(t_w)$ as a function of aging time, $t_w$, for {Laponite} clay suspensions of concentration, $C_L = $ 2.5\% w/v prepared in pure water or in NaCl or TSPP solutions. Error bars are standard errors from three independent measurements. (c) $Z_{eff}(t_w\approx0)$
     \textit{vs.} standard deviation of surface intensity in arbitrary units. (Inset) $Z_{eff}(t_w\approx0)$ on the Latex microsphere trapped in water or clay suspensions.}
     \label{fig:6}
 \end{figure*}
Figure~\ref{fig:6} displays the results {when the microsphere was trapped in Laponite} clay suspensions of concentration 2.5\% w/v with and without externally added salts. 
{E}nhanced screening of {the} electrostatic repulsions between the microsphere and Laponite faces in the presence of NaCl results in very rapid increase in $Z_{eff}$ (Fig.~\ref{fig:6}(a)), and therefore, also of $\gamma$ ({Fig.~\ref{fig:6}(b)}). We attribute this observation to the accelerated adsorption of single Laponite nanoplatelets, clay aggregates and even gel network strands {(Fig.~\ref{fig:3-2}(a))} on the microsphere~\cite{C2SM25731A}. {T}he very large error bars that are seen for higher $t_w$, $C_L$ and NaCl concentration imply increased heterogeneity in the adsorption process. 

In the presence of TSPP, $Z_{eff}$ does not show noticeable variation as a function of $t_w$. We note {from Fig.~\ref{fig:6}(a)} that  $Z_{eff}(t_{w}{\approx}0)$ in the presence of TSPP is higher than that {for} water. Since TSPP is expected to render negative charges on the rims of the nanoplatelets~\cite{PhysRevE.66.021401}, electrostatically-driven adsorption on the weakly negatively charged microsphere can be ruled out. {We conclude, t}herefore, {that} the adsorption process in this scenario is driven completely by non-electrostatic dispersion forces. In contrast, dispersion forces {and electrostatic attractions} are both important in the presence of NaCl. The weakened electrostatic repulsion between like-charged surfaces results in accelerated adsorption in an ionic medium, as seen from Fig.~\ref{fig:6}(a). {A similar} mechanism was proposed to explain the adsorption of Laponite nanoplatelet (Grade: RDS) on polystyrene particles during emulsion polymerization~\cite{doi:10.1021/acs.langmuir.5b03576}.
\begin{figure}[t]
     \centering
     \includegraphics[width=1\linewidth]{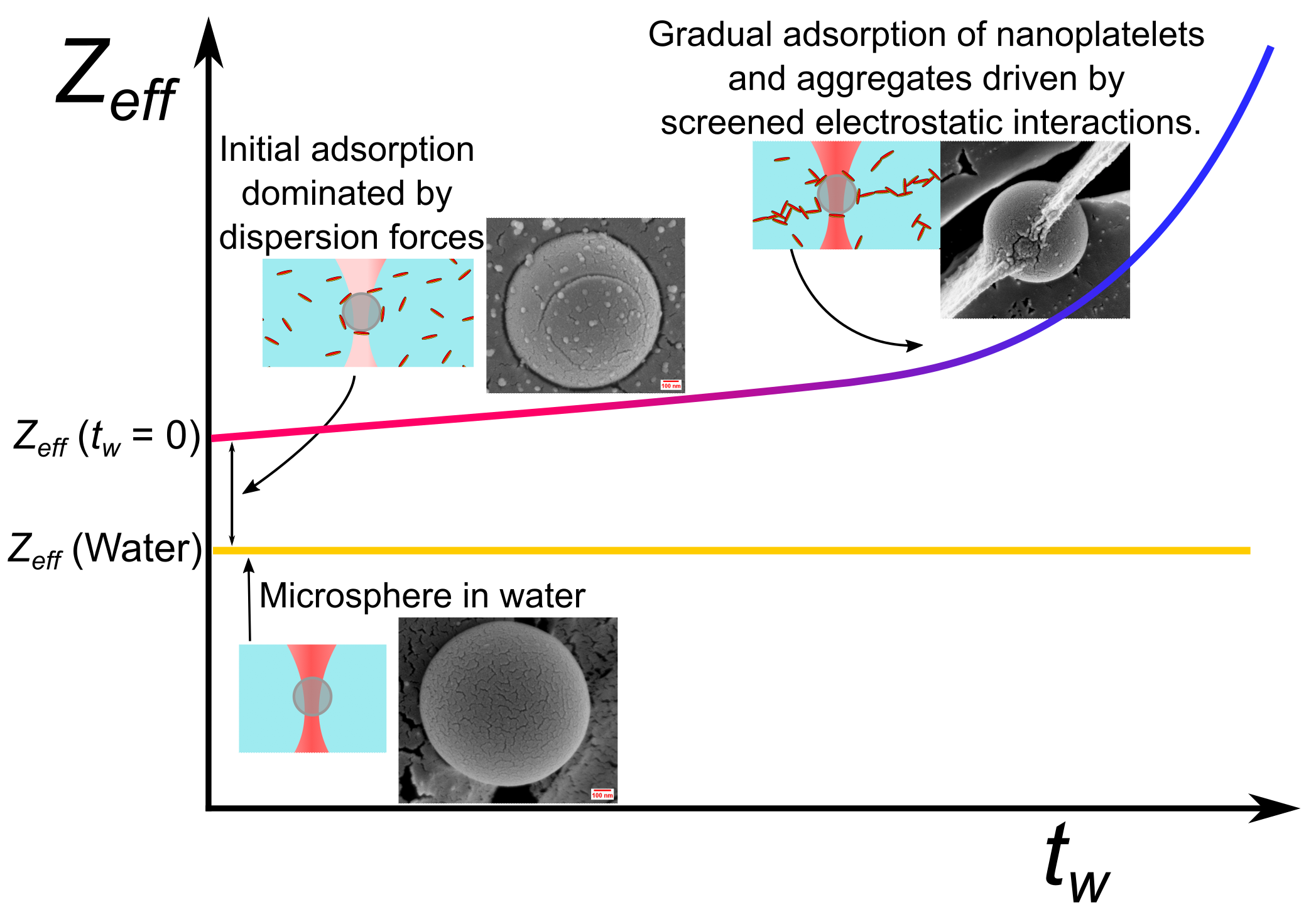}
     \caption{Schematic representation of {the} mechanisms governing the adsorption process of clay nanoplatelets on a Latex microsphere.}
     \label{fig:7}
 \end{figure}

We have displayed $Z_{eff}(t_w \approx 0)$ ({i}nset of Fig.~\ref{fig:6}(c){, estimated from data presented in Fig.~\ref{fig:6}(a)}) as a function of the standard deviation of microsphere surface intensities (Fig.~\ref{fig:3-2}(b)) in Figure~\ref{fig:6}{(c)}. 
{The} direct correlation observed between $Z_{eff}(t_w \approx 0)$ and {the} standard deviation of surface intensit{ies} suggests a qualitative agreement between our analyses of {clay nanoplatelet} adsorption {on a Latex microsphere} using single-colloid electrophoresis and cryo-FESEM imaging. Finally, we summarize {the} insights {gained }on the adsorption process of clay nanoplatelets on a Latex microsphere in Fig.~\ref{fig:7}.  {T}he adsorption that {initially} occurs is dominated by non-electrostatic dispersion interactions. {When the microsphere was trapped in} gel-forming {clay} samples above a critical concentration, the rapidly increasing $Z_{eff}(t_w)$ {at later times} suggests gradual adsorption of clay nanoplatelets, clay aggregates and gel network strands onto the microsphere {surface, emphasising, thereby, }the important role of {screened electrostatic interactions}.

	\section{\label{se:sac} Conclusions}
	In this report, we investigated, for the first time to the best of our knowledge, the adsorption of charged colloidal clay {Laponite} nanoplatelets on a preformed Latex microsphere in an aqueous medium. {In} a medium of pH < 11, a Laponite clay nanoplatelet has negatively charged faces and weakly positively charged rims~\cite{TAWARI200154}. Above a critical concentration of 1\% w/v, these nanoplatelets self-assemble gradually to form system-spanning gel networks which reform continuously due to nanoplatelet rearrangements in a physical aging process. {In the present research, we first visualised the adsorption of Laponite clay nanoplatelets on a Latex microsphere using field emission scanning electron microscopy and next quantified the kinetics of the adsorption process} by performing optical tweezer-based single-colloid electrophoresis measurements. 

Complex potentials can be created in optical experiments, which can be employed to study and control colloidal interactions and dynamics~\cite{D4SM00853G} and to verify postulates of non-equilibrium statistical mechanics~\cite{PhysRevX.7.021051}. In the present work, optical tweezers were used to perform sensitive measurements of {the} effective surface charge and Stokes drag on optically trapped {Latex} microsphere{s} {in aqueous suspensions of charged Laponite clay nanoplatelets in the presence of an alternating electric field.} An increase in the clay nanoplatelet concentration in the suspension increased the {measured} effective surface charges on the trapped microsphere due to enhanced nanoplatelet adsorption. The initial adsorption process was very fast and predominantly driven by non-electrostatic dispersion forces. When the microsphere was trapped in clay suspensions of concentration{s} 2.0 - 2.8\% w/v, {we} observed {a nonlinear} increase {in} the effective surface charge {at later times} due to weakening of electrostatic repulsive interactions between the nanoplatelet faces and the surface of the trapped microsphere. 
{Finally, w}e reported a very good agreement between the results obtained {from} cryo-FESEM and optical tweezer-based single-colloid electrophoresis.

Our work establishes optical tweezer-based single-colloid electrophoresis as a technique that has enormous potential in the study of adsorption of charged colloids on solid surfaces. This technique can be implemented successfully in applications involving colloidal stabilisation, for example, while preparing clay-armoured particles {for} surfactant-free Pickering emulsion polymerisation, and as a sensor for label-free detection in biomedical applications. The ability to measure small variations in the charges on a solid surface can be exploited to detect the adsorption of biomolecules. A combination of microfluidic flow cells and high{er} electric field strengths can be employed in future studies to attain {even} better signal-to-noise ratio{s. {This} would significantly} enhance temporal resolution and efficiency {in the detection of} macromolecular adsorption events.


\section*{Author contributions}
\textbf{Vaibhav Raj Singh Parmar:} Methodology, Formal analysis, Investigation, Conceptualization, Writing – original draft, Software, Visualization, Writing – review \& editing, Data curation. \textbf{Sayantan Chanda:} Data curation, Formal analysis, Investigation, Validation, Software, Writing – original draft, Writing – review \& editing. \textbf{Sri Vishnu Bharat S:} Data curation, Validation, Investigation, Writing – original draft, Writing – review \& editing. \textbf{Ranjini Bandyopadhyay:} Conceptualization, Funding acquisition, Project administration, Resources, Supervision, Visualization, Writing – review \& editing.
	
	\section*{Conflicts of interest}
There are no conflicts to declare.

\section*{Data availability}
The analysis codes and all the data sets related to this article are deposited in the GitHub repository (\href{https://github.com/vaibhav00parmar/Optical-tweezer-electrophoresis-clay-nanoplatelet-adsorption-on-Latex-microspheres-in-aqueous-media.git}{https://github.com/vaibhav00parmar/Optical-tweezer-electrophoresis-clay-nanoplatelet-adsorption-on-Latex-microspheres-in-aqueous-media.git}).

\section*{Acknowledgements}
The authors acknowledge help from Mr. K. M. Yatheendran in acquiring cryo-FESEM images and Dr. Rajkumar Biswas in fabricating the sample cells during the preliminary stage of the experiment.

\renewcommand\refname{References}	
\bibliographystyle{elsarticle-num}
 \bibliography{rsc}

\begin{thebibliography}{10}
\expandafter\ifx\csname url\endcsname\relax
  \def\url#1{\texttt{#1}}\fi
\expandafter\ifx\csname urlprefix\endcsname\relax\def\urlprefix{URL }\fi
\expandafter\ifx\csname href\endcsname\relax
  \def\href#1#2{#2} \def\path#1{#1}\fi

\bibitem{ZHANG200035}
Y.~Zhang, N.~W. Franklin, R.~J. Chen, H.~Dai, Metal coating on suspended carbon nanotubes and its implication to metal–tube interaction, Chem. Phys. Lett. 331~(1) (2000) 35--41.
\newblock \href {https://doi.org/https://doi.org/10.1016/S0009-2614(00)01162-3} {\path{doi:https://doi.org/10.1016/S0009-2614(00)01162-3}}.

\bibitem{doi:10.1021/la502588g}
G.~Marzun, C.~Streich, S.~Jendrzej, S.~Barcikowski, P.~Wagener, Adsorption of colloidal platinum nanoparticles to supports: Charge transfer and effects of electrostatic and steric interactions, Langmuir 30~(40) (2014) 11928--11936.
\newblock \href {https://doi.org/10.1021/la502588g} {\path{doi:10.1021/la502588g}}.

\bibitem{doi:10.1126/science.1135752}
D.~Matthey, J.~G. Wang, S.~Wendt, J.~Matthiesen, R.~Schaub, E.~Lægsgaard, B.~Hammer, F.~Besenbacher, \href{https://www.science.org/doi/abs/10.1126/science.1135752}{Enhanced bonding of gold nanoparticles on oxidized tio<sub>2</sub>(110)}, Science 315~(5819) (2007) 1692--1696.
\newblock \href {https://doi.org/10.1126/science.1135752} {\path{doi:10.1126/science.1135752}}.
\newline\urlprefix\url{https://www.science.org/doi/abs/10.1126/science.1135752}

\bibitem{Adeyemo2017}
A.~A. Adeyemo, I.~O. Adeoye, O.~S. Bello, \href{https://doi.org/10.1007/s13201-015-0322-y}{{Adsorption of dyes using different types of clay: a review}}, Appl. Water Sci. 7~(2) (2017) 543--568.
\newblock \href {https://doi.org/10.1007/s13201-015-0322-y} {\path{doi:10.1007/s13201-015-0322-y}}.
\newline\urlprefix\url{https://doi.org/10.1007/s13201-015-0322-y}

\bibitem{doi:10.1021/acs.langmuir.5b03576}
B.~Brunier, N.~Sheibat-Othman, Y.~Chevalier, E.~Bourgeat-Lami, \href{https://doi.org/10.1021/acs.langmuir.5b03576}{{Partitioning of Laponite Clay Platelets in Pickering Emulsion Polymerization}}, Langmuir 32~(1) (2016) 112--124, pMID: 26653971.
\newblock \href {https://doi.org/10.1021/acs.langmuir.5b03576} {\path{doi:10.1021/acs.langmuir.5b03576}}.
\newline\urlprefix\url{https://doi.org/10.1021/acs.langmuir.5b03576}

\bibitem{GICHEVA201351}
G.~Gicheva, G.~Yordanov, Removal of citrate-coated silver nanoparticles from aqueous dispersions by using activated carbon, Colloids Surf. A Physicochem. Eng. Asp. 431 (2013) 51--59.
\newblock \href {https://doi.org/https://doi.org/10.1016/j.colsurfa.2013.04.039} {\path{doi:https://doi.org/10.1016/j.colsurfa.2013.04.039}}.

\bibitem{doi:https://doi.org/10.1002/9781118881194.ch7}
G.~Kontogeorgis, S.~Kiil, Introduction to Applied Colloid and Surface Chemistry (eds G.M. Kontogeorgis and S. Kiil), John Wiley \& Sons, Ltd, 2016, Ch.~7, pp. 161--184.
\newblock \href {https://doi.org/https://doi.org/10.1002/9781118881194.ch7} {\path{doi:https://doi.org/10.1002/9781118881194.ch7}}.

\bibitem{doi:10.1021/acs.langmuir.6b01080}
B.~Brunier, N.~Sheibat-Othman, M.~Chniguir, Y.~Chevalier, E.~Bourgeat-Lami, Investigation of four different laponite clays as stabilizers in pickering emulsion polymerization, Langmuir 32~(24) (2016) 6046--6057.
\newblock \href {https://doi.org/10.1021/acs.langmuir.6b01080} {\path{doi:10.1021/acs.langmuir.6b01080}}.

\bibitem{Teixeira2011}
R.~F.~A. Teixeira, H.~S. McKenzie, A.~A. Boyd, S.~A.~F. Bon, \href{https://doi.org/10.1021/ma201691u}{{Pickering Emulsion Polymerization Using Laponite Clay as Stabilizer To Prepare Armored ``Soft'' Polymer Latexes}}, Macromolecules 44~(18) (2011) 7415--7422.
\newblock \href {https://doi.org/10.1021/ma201691u} {\path{doi:10.1021/ma201691u}}.
\newline\urlprefix\url{https://doi.org/10.1021/ma201691u}

\bibitem{B904740A}
T.~Wang, P.~J. Colver, S.~A.~F. Bon, J.~L. Keddie, \href{http://dx.doi.org/10.1039/B904740A}{{Soft polymer and nano-clay supracolloidal particles in adhesives: synergistic effects on mechanical properties}}, Soft Matter 5 (2009) 3842--3849.
\newblock \href {https://doi.org/10.1039/B904740A} {\path{doi:10.1039/B904740A}}.
\newline\urlprefix\url{http://dx.doi.org/10.1039/B904740A}

\bibitem{claybook}
H.~Van~Olphen, {An Introduction to Clay Colloid Chemistry: For Clay Technologists, Geologists and Soil Scientists. 2nd edition}, Wiley, New York, 1977.

\bibitem{THOMPSON1992236}
D.~W. Thompson, J.~T. Butterworth, \href{https://www.sciencedirect.com/science/article/pii/002197979290254J}{The nature of laponite and its aqueous dispersions}, J. Colloid Interface Sci. 151~(1) (1992) 236--243.
\newblock \href {https://doi.org/https://doi.org/10.1016/0021-9797(92)90254-J} {\path{doi:https://doi.org/10.1016/0021-9797(92)90254-J}}.
\newline\urlprefix\url{https://www.sciencedirect.com/science/article/pii/002197979290254J}

\bibitem{doi:10.1021/acs.langmuir.8b01830}
K.~Suman, Y.~M. Joshi, \href{https://doi.org/10.1021/acs.langmuir.8b01830}{{Microstructure and Soft Glassy Dynamics of an Aqueous Laponite Dispersion}}, Langmuir 34~(44) (2018) 13079--13103, pMID: 30180583.
\newblock \href {https://doi.org/10.1021/acs.langmuir.8b01830} {\path{doi:10.1021/acs.langmuir.8b01830}}.
\newline\urlprefix\url{https://doi.org/10.1021/acs.langmuir.8b01830}

\bibitem{doi:10.1021/acs.langmuir.5b00291}
D.~Saha, R.~Bandyopadhyay, Y.~M. Joshi, {Dynamic Light Scattering Study and DLVO Analysis of Physicochemical Interactions in Colloidal Suspensions of Charged Disks}, Langmuir 31~(10) (2015) 3012--3020.
\newblock \href {https://doi.org/10.1021/acs.langmuir.5b00291} {\path{doi:10.1021/acs.langmuir.5b00291}}.

\bibitem{Ruzicka2011}
B.~Ruzicka, E.~Zaccarelli, L.~Zulian, R.~Angelini, M.~Sztucki, A.~Moussa{\"i}d, T.~Narayanan, F.~Sciortino, \href{https://doi.org/10.1038/nmat2921}{Observation of empty liquids and equilibrium gels in a colloidal clay}, Nature Materials 10~(1) (2011) 56--60.
\newblock \href {https://doi.org/10.1038/nmat2921} {\path{doi:10.1038/nmat2921}}.
\newline\urlprefix\url{https://doi.org/10.1038/nmat2921}

\bibitem{C0SM00590H}
B.~Ruzicka, E.~Zaccarelli, \href{http://dx.doi.org/10.1039/C0SM00590H}{{A fresh look at the Laponite phase diagram}}, Soft Matter 7 (2011) 1268--1286.
\newblock \href {https://doi.org/10.1039/C0SM00590H} {\path{doi:10.1039/C0SM00590H}}.
\newline\urlprefix\url{http://dx.doi.org/10.1039/C0SM00590H}

\bibitem{C2SM25731A}
M.~Delhorme, B.~Jönsson, C.~Labbez, Monte carlo simulations of a clay inspired model suspension: the role of rim charge, Soft Matter 8 (2012) 9691--9704.
\newblock \href {https://doi.org/10.1039/C2SM25731A} {\path{doi:10.1039/C2SM25731A}}.

\bibitem{D2SM01457B}
R.~Biswas, V.~R.~S. Parmar, A.~G. Thambi, R.~Bandyopadhyay, Correlating microscopic viscoelasticity and structure of an aging colloidal gel using active microrheology and cryogenic scanning electron microscopy, Soft Matter 19 (2023) 2407--2416.
\newblock \href {https://doi.org/10.1039/D2SM01457B} {\path{doi:10.1039/D2SM01457B}}.

\bibitem{PhysRevLett.93.228302}
R.~Bandyopadhyay, D.~Liang, H.~Yardimci, D.~A. Sessoms, M.~A. Borthwick, S.~G.~J. Mochrie, J.~L. Harden, R.~L. Leheny, {Evolution of Particle-Scale Dynamics in an Aging Clay Suspension}, Phys. Rev. Lett. 93 (2004) 228302.
\newblock \href {https://doi.org/10.1103/PhysRevLett.93.228302} {\path{doi:10.1103/PhysRevLett.93.228302}}.

\bibitem{D1SM00987G}
C.~Misra, V.~T. Ranganathan, R.~Bandyopadhyay, {Influence of medium structure on the physicochemical properties of aging colloidal dispersions investigated using the synthetic clay LAPONITE}, Soft Matter 17 (2021) 9387--9398.
\newblock \href {https://doi.org/10.1039/D1SM00987G} {\path{doi:10.1039/D1SM00987G}}.

\bibitem{MONGONDRY2004191}
P.~Mongondry, T.~Nicolai, J.-F. Tassin, \href{https://www.sciencedirect.com/science/article/pii/S0021979704000955}{Influence of pyrophosphate or polyethylene oxide on the aggregation and gelation of aqueous laponite dispersions}, J. Colloid Interface Sci. 275~(1) (2004) 191--196.
\newblock \href {https://doi.org/https://doi.org/10.1016/j.jcis.2004.01.037} {\path{doi:https://doi.org/10.1016/j.jcis.2004.01.037}}.
\newline\urlprefix\url{https://www.sciencedirect.com/science/article/pii/S0021979704000955}

\bibitem{PhysRevE.66.021401}
C.~Martin, F.~Pignon, J.-M. Piau, A.~Magnin, P.~Lindner, B.~Cabane, Dissociation of thixotropic clay gels, Phys. Rev. E 66 (2002) 021401.
\newblock \href {https://doi.org/10.1103/PhysRevE.66.021401} {\path{doi:10.1103/PhysRevE.66.021401}}.

\bibitem{doi:10.1021/la980117p}
A.~Mourchid, E.~Lécolier, H.~Van~Damme, P.~Levitz, \href{https://doi.org/10.1021/la980117p}{{On Viscoelastic, Birefringent, and Swelling Properties of Laponite Clay Suspensions: Revisited Phase Diagram}}, Langmuir 14~(17) (1998) 4718--4723.
\newblock \href {https://doi.org/10.1021/la980117p} {\path{doi:10.1021/la980117p}}.
\newline\urlprefix\url{https://doi.org/10.1021/la980117p}

\bibitem{GARBOW2001227}
N.~Garbow, M.~Evers, T.~Palberg, \href{https://www.sciencedirect.com/science/article/pii/S0927775701008469}{Optical tweezing electrophoresis of isolated, highly charged colloidal spheres}, Colloids Surf. A Physicochem. Eng. Asp. 195~(1) (2001) 227--241.
\newblock \href {https://doi.org/https://doi.org/10.1016/S0927-7757(01)00846-9} {\path{doi:https://doi.org/10.1016/S0927-7757(01)00846-9}}.
\newline\urlprefix\url{https://www.sciencedirect.com/science/article/pii/S0927775701008469}

\bibitem{10.1063/1.2884147}
O.~Otto, C.~Gutsche, F.~Kremer, U.~F. Keyser, {Optical tweezers with 2.5kHz bandwidth video detection for single-colloid electrophoresis}, Rev. Sci. Instrum. 79~(2) (2008) 023710.
\newblock \href {https://doi.org/10.1063/1.2884147} {\path{doi:10.1063/1.2884147}}.

\bibitem{PESCE2014568}
G.~Pesce, G.~Rusciano, A.~Sasso, R.~Isticato, T.~Sirec, E.~Ricca, \href{https://www.sciencedirect.com/science/article/pii/S092777651400040X}{Surface charge and hydrodynamic coefficient measurements of bacillus subtilis spore by optical tweezers}, Colloids Surf. B 116 (2014) 568--575.
\newblock \href {https://doi.org/https://doi.org/10.1016/j.colsurfb.2014.01.039} {\path{doi:https://doi.org/10.1016/j.colsurfb.2014.01.039}}.
\newline\urlprefix\url{https://www.sciencedirect.com/science/article/pii/S092777651400040X}

\bibitem{10.1063/1.4967401}
G.~Kokot, M.~I. Bespalova, M.~Krishnan, {Measured electrical charge of SiO2 in polar and nonpolar media}, J. Chem. Phys. 145~(19) (2016) 194701.
\newblock \href {https://doi.org/10.1063/1.4967401} {\path{doi:10.1063/1.4967401}}.

\bibitem{10.1063/1.2734968}
G.~Seth~Roberts, T.~A. Wood, W.~J. Frith, P.~Bartlett, {Direct measurement of the effective charge in nonpolar suspensions by optical tracking of single particles}, J. Chem. Phys. 126~(19) (2007) 194503.
\newblock \href {https://doi.org/10.1063/1.2734968} {\path{doi:10.1063/1.2734968}}.

\bibitem{PhysRevLett.108.016101}
F.~Beunis, F.~Strubbe, K.~Neyts, D.~Petrov, \href{https://link.aps.org/doi/10.1103/PhysRevLett.108.016101}{Beyond millikan: The dynamics of charging events on individual colloidal particles}, Phys. Rev. Lett. 108 (2012) 016101.
\newblock \href {https://doi.org/10.1103/PhysRevLett.108.016101} {\path{doi:10.1103/PhysRevLett.108.016101}}.
\newline\urlprefix\url{https://link.aps.org/doi/10.1103/PhysRevLett.108.016101}

\bibitem{doi:10.1021/acsnano.3c08161}
Y.~Ussembayev, F.~Beunis, L.~Oorlynck, M.~Bahrami, F.~Strubbe, K.~Neyts, Single elementary charge fluctuations on nanoparticles in aqueous solution, ACS Nano 17~(22) (2023) 22952--22959.
\newblock \href {https://doi.org/10.1021/acsnano.3c08161} {\path{doi:10.1021/acsnano.3c08161}}.

\bibitem{Grollman:17}
R.~Grollman, G.~Founds, R.~Wallace, O.~Ostroverkhova, \href{https://opg.optica.org/oe/abstract.cfm?URI=oe-25-23-29161}{Simultaneous fluorescence and surface charge measurements on organic semiconductor-coated silica microspheres in (non)polar liquids}, Opt. Express 25~(23) (2017) 29161--29171.
\newblock \href {https://doi.org/10.1364/OE.25.029161} {\path{doi:10.1364/OE.25.029161}}.
\newline\urlprefix\url{https://opg.optica.org/oe/abstract.cfm?URI=oe-25-23-29161}

\bibitem{10.1063/1.4922039}
T.~Brans, F.~Strubbe, C.~Schreuer, K.~Neyts, F.~Beunis, {Optical tweezing electrophoresis of single biotinylated colloidal particles for avidin concentration measurement}, J. Appl. Phys. 117~(21) (2015) 214704.
\newblock \href {https://doi.org/10.1063/1.4922039} {\path{doi:10.1063/1.4922039}}.

\bibitem{OKADA200817}
T.~Okada, Y.~Yamamoto, T.~Shibuya, H.-W. Kang, H.~Miyachi, I.~Karube, H.~Muramatsu, J.~M. Kim, \href{https://www.sciencedirect.com/science/article/pii/S1369703X0800082X}{{Development of AC microelectrophoresis for rapid protein affinity evaluation}}, Biochem. Eng. J. 41~(1) (2008) 17--23.
\newblock \href {https://doi.org/https://doi.org/10.1016/j.bej.2008.03.001} {\path{doi:https://doi.org/10.1016/j.bej.2008.03.001}}.
\newline\urlprefix\url{https://www.sciencedirect.com/science/article/pii/S1369703X0800082X}

\bibitem{GEONZON2023846}
L.~C. Geonzon, M.~Kobayashi, T.~Sugimoto, Y.~Adachi, \href{https://www.sciencedirect.com/science/article/pii/S0021979722018288}{Adsorption kinetics of polyacrylamide-based polyelectrolyte onto a single silica particle studied using microfluidics and optical tweezers}, J. Colloid Interface Sci. 630 (2023) 846--854.
\newblock \href {https://doi.org/https://doi.org/10.1016/j.jcis.2022.10.067} {\path{doi:https://doi.org/10.1016/j.jcis.2022.10.067}}.
\newline\urlprefix\url{https://www.sciencedirect.com/science/article/pii/S0021979722018288}

\bibitem{C005217P}
J.~A. van Heiningen, R.~J. Hill, Polymer adsorption onto a micro-sphere from optical tweezers electrophoresis, Lab Chip 11 (2011) 152--162.
\newblock \href {https://doi.org/10.1039/C005217P} {\path{doi:10.1039/C005217P}}.

\bibitem{C0CP02296A}
E.~Bianchi, R.~Blaak, C.~N. Likos, Patchy colloids: state of the art and perspectives, Phys. Chem. Chem. Phys. 13 (2011) 6397--6410.
\newblock \href {https://doi.org/10.1039/C0CP02296A} {\path{doi:10.1039/C0CP02296A}}.

\bibitem{10.1063/1.2356852}
S.~F. Tolić-Nørrelykke, E.~Schäffer, J.~Howard, F.~S. Pavone, F.~Jülicher, H.~Flyvbjerg, {Calibration of optical tweezers with positional detection in the back focal plane}, Rev. Sci. Instrum. 77~(10) (2006) 103101.
\newblock \href {https://doi.org/10.1063/1.2356852} {\path{doi:10.1063/1.2356852}}.

\bibitem{10.1063/1.1785844}
K.~C. Neuman, S.~M. Block, {Optical trapping}, Rev. Sci. Instrum. 75~(9) (2004) 2787--2809.
\newblock \href {https://doi.org/10.1063/1.1785844} {\path{doi:10.1063/1.1785844}}.

\bibitem{B703994H}
T.~A. Wood, G.~S. Roberts, S.~Eaimkhong, P.~Bartlett, Characterization of microparticles with driven optical tweezers, Faraday Discuss. 137 (2008) 319--333.
\newblock \href {https://doi.org/10.1039/B703994H} {\path{doi:10.1039/B703994H}}.

\bibitem{SEMENOV2009260}
I.~Semenov, O.~Otto, G.~Stober, P.~Papadopoulos, U.~Keyser, F.~Kremer, \href{https://www.sciencedirect.com/science/article/pii/S0021979709006602}{Single colloid electrophoresis}, J. Colloid Interface Sci. 337~(1) (2009) 260--264.
\newblock \href {https://doi.org/https://doi.org/10.1016/j.jcis.2009.05.017} {\path{doi:https://doi.org/10.1016/j.jcis.2009.05.017}}.
\newline\urlprefix\url{https://www.sciencedirect.com/science/article/pii/S0021979709006602}

\bibitem{10.1002/elps.201300214}
G.~Pesce, V.~Lisbino, G.~Rusciano, A.~Sasso, Optical manipulation of charged microparticles in polar fluids, Electrophoresis 34~(22-23) (2013) 3141--3149.
\newblock \href {https://doi.org/https://doi.org/10.1002/elps.201300214} {\path{doi:https://doi.org/10.1002/elps.201300214}}.

\bibitem{2024arXiv240701396S}
V.~R. {Singh Parmar}, R.~{Bandyopadhyay}, {Manipulating crack formation in air-dried clay suspensions with tunable elasticity}, arXiv e-prints (2024) arXiv:2407.01396\href {http://arxiv.org/abs/2407.01396} {\path{arXiv:2407.01396}}, \href {https://doi.org/10.48550/arXiv.2407.01396} {\path{doi:10.48550/arXiv.2407.01396}}.

\bibitem{RAVIKUMAR2008326}
N.~{Ravi Kumar}, K.~Muralidhar, Y.~M. Joshi, {On the refractive index of ageing dispersions of Laponite}, Appl. Clay. Sci. 42~(1) (2008) 326--330.
\newblock \href {https://doi.org/https://doi.org/10.1016/j.clay.2007.12.010} {\path{doi:https://doi.org/10.1016/j.clay.2007.12.010}}.

\bibitem{10.1063/1.480582}
S.~Kutter, J.-P. Hansen, M.~Sprik, E.~Boek, \href{https://doi.org/10.1063/1.480582}{{Structure and phase behavior of a model clay dispersion: A molecular-dynamics investigation}}, J. Chem. Phys. 112~(1) (2000) 311--322.
\newblock \href {https://doi.org/10.1063/1.480582} {\path{doi:10.1063/1.480582}}.
\newline\urlprefix\url{https://doi.org/10.1063/1.480582}

\bibitem{TAWARI200154}
S.~L. Tawari, D.~L. Koch, C.~Cohen, Electrical double-layer effects on the brownian diffusivity and aggregation rate of laponite clay particles, J. Colloid Interface Sci. 240~(1) (2001) 54--66.
\newblock \href {https://doi.org/https://doi.org/10.1006/jcis.2001.7646} {\path{doi:https://doi.org/10.1006/jcis.2001.7646}}.

\bibitem{D4SM00853G}
D.~Saha, S.~Tarama, H.~Löwen, S.~U. Egelhaaf, \href{http://dx.doi.org/10.1039/D4SM00853G}{Cybloids – creation and control of cybernetic colloids}, Soft Matter 20 (2024) 8112--8124.
\newblock \href {https://doi.org/10.1039/D4SM00853G} {\path{doi:10.1039/D4SM00853G}}.
\newline\urlprefix\url{http://dx.doi.org/10.1039/D4SM00853G}

\bibitem{PhysRevX.7.021051}
S.~Ciliberto, \href{https://link.aps.org/doi/10.1103/PhysRevX.7.021051}{Experiments in stochastic thermodynamics: Short history and perspectives}, Phys. Rev. X 7 (2017) 021051.
\newblock \href {https://doi.org/10.1103/PhysRevX.7.021051} {\path{doi:10.1103/PhysRevX.7.021051}}.
\newline\urlprefix\url{https://link.aps.org/doi/10.1103/PhysRevX.7.021051}

\end{thebibliography}

 \pagebreak
\renewcommand{\figurename}{Supplementary Fig.}
\begin{center}
	\textbf{\LARGE Supplementary Material}\\
	\vspace{0.4cm}
	\textbf{\LARGE \textcolor{red}{Using optical tweezer electrophoresis to investigate clay nanoplatelet adsorption on Latex microsphere{s} in aqueous media}}\\
	\vspace{0.4cm}
	{Vaibhav Raj Singh Parmar}, {Sayantan Chanda}, 
 {Sri Vishnu Bharat Sivasubramaniam} and {Ranjini Bandyopadhyay}$^*$\\
	\vspace{0.4cm}
	\textit{Soft Condensed Matter Group, Raman Research Institute, C. V. Raman Avenue, Sadashivanagar, Bangalore 560 080, INDIA}
	
	\date{\today}
\end{center}

\footnotetext[1]{Corresponding Author: Ranjini Bandyopadhyay; Email: ranjini@rri.res.in}
\maketitle

\definecolor{black}{rgb}{0.0, 0.0, 0.0}
\definecolor{red(ryb)}{rgb}{1.0, 0.15, 0.07}
\definecolor{darkred}{rgb}{0.55, 0.0, 0.0}
\definecolor{blue(ryb)}{rgb}{0.01, 0.2, 1.0}
\definecolor{darkcyan}{rgb}{0.0, 0.55, 0.55}
\definecolor{navyblue}{rgb}{0.0, 0.0, 0.5}
\definecolor{olivedrab(web)(olivedrab3)}{rgb}{0.42, 0.56, 0.14}
\definecolor{darkraspberry}{rgb}{0.50, 0.0, 1.0}
\definecolor{magenta}{rgb}{1.0, 0.0, 1.0}
\newcommand{\pcircle}{\textcolor{darkraspberry}{\large$\bullet$}}
\newcommand{\phex}{\textcolor{darkraspberry}{\large$\varhexagonblack$}}
\newcommand{\rdhexagon}{\textcolor{red(ryb)}{\small$\blackdiamond$}}

\setcounter{table}{0}
\renewcommand{\thetable}{S\arabic{table}}%
\renewcommand{\tablename}{Supplementary Table}
\setcounter{figure}{0}
\makeatletter 
\renewcommand{\figurename}{Supplementary Fig.}
\setcounter{figure}{0}
\makeatletter 
\renewcommand{\thefigure}{S\arabic{figure}}
\setcounter{section}{0}
\renewcommand{\thesection}{ST\arabic{section}}
\setcounter{equation}{0}
\renewcommand{\theequation}{S\arabic{equation}}

	\section{Cryogenic field emission scanning electron microscopy (cryo-FESEM) images of Latex microsphere}

 \begin{figure}[H]
     \centering
     \includegraphics[width=1.0\linewidth]{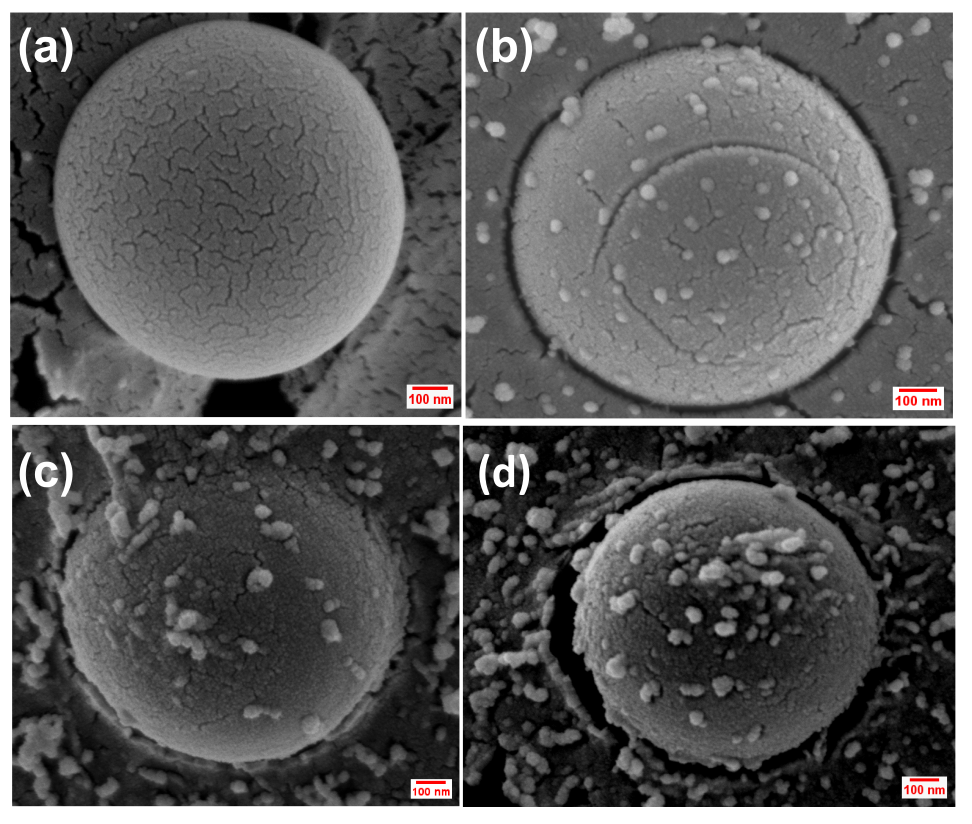}
     \caption{Raw cryo-FESEM images of {a trapped} Latex microsphere in {(a)} pure water, {(b)} {an} aqueous Laponite suspension of concentration 2.5\% w/v, {(c)} an aqueous Laponite suspension with 1.25 mM NaCl and {(d)} {an aqueous} Laponite suspension with 1.25 mM TSPP. 
     }
     \label{fig:yu}
 \end{figure}
 
 \begin{figure}[H]
	 	\includegraphics[width= 6.0in]{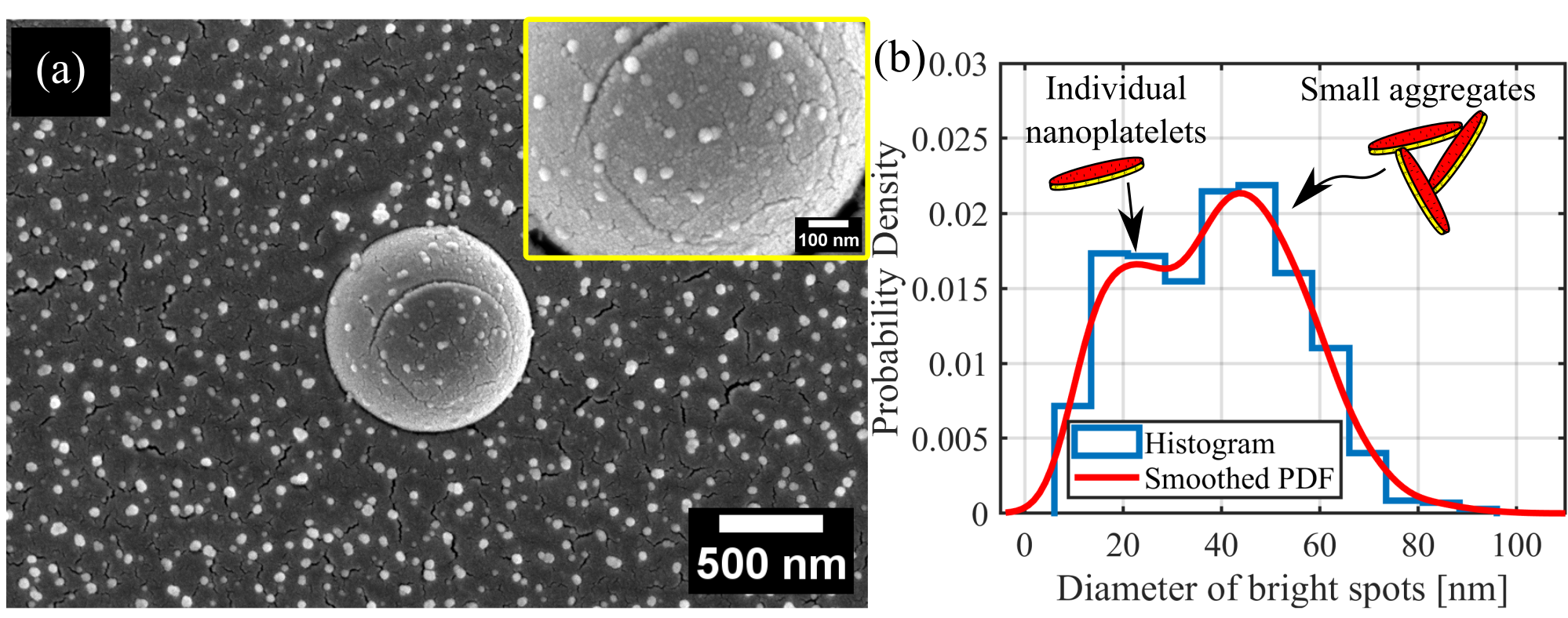}
	 	\centering
	 	\caption{\label{SI1} {(a)} Cryo-{FE}SEM image of a Latex microsphere (diameter = $1$ $\mu$m) suspended in a clay suspension of concentration 2.5\% w/v at aging time $t_w = $ 90 minutes. (Inset) Higher magnification image of the Latex microsphere surface. Many bright spots can be seen on the surface of the microsphere. {(b)} Histogram and smoothed probability density of the diameters of 932 bright spots identified from (a) using ImageJ {and} estimated using the histogram and ksdensity functions in MATLAB@2024. 
   The diameter of each bright spot was estimated from its area, $A_{spot}$, 
   by calculating {2${\sqrt{{A_{{spot}}/\pi}}}$}. The peak in the probability density function at $\sim$ 25 nm corresponds to an individual clay nanoplatelet{, while the second peak at a larger diameter corresponds} to small clay aggregates}.
\end{figure}


\newpage
 \section{Calibrations of the optical tweezer}
 \subsection{Measurement of QPD sensitivity}
 The QPD sensitivity, $\beta$, determined from the uncalibrated power spectrum, $S_{vv}$, of the trapped microsphere {in QPD units}, is related to the position of the microsphere as follows: $x =\beta{x_{v}}$, where $x_{v}$ is the position of the microsphere in QPD units. The power spectrum of the particle is given by:
 \begin{equation}
     S_{xx}=\int_{-\infty}^{\infty}e^{-i\omega t} \int_{-\infty}^{\infty}x(t)x(t+\tau) \ d\tau\ dt
 \end{equation}
 
{Therefore,} $S_{xx}=\beta^2 S_{vv}$.
 For a trapped Brownian particle, $S_{xx}$ is theoretically calculated from the first term in Eqn.~({2}) of the main manuscript:
\begin{equation}
    S_{xx} = \beta^2S_{vv}=\frac{ k_B T}{\pi^2\gamma (f_c^2+f^2)}
\end{equation}

 For frequencies $f>>f_c$,
 \begin{equation}
     f^2S_{vv}=\beta^{-2}\frac{k_BT}{\pi^2\gamma}
 \end{equation}

This indicates that $f^2S_{vv}$ attains a constant value at high frequencies. For a microsphere trapped in water, the plateau is clearly visible in Fig.~S{3}(a). Since the {Stokes drag coefficient} $\gamma = 6\pi \eta a$, where $a$ is the radius of the microsphere, $\eta$ is the medium viscosity and $T$ is temperature of the medium, is known for a microsphere in water, QPD sensitivity, $\beta$, can be determined using Eqn.~(S3). This method is less prone to errors as the possibility of experimental noise at lower frequencies is eliminated. QPD sensitivity, $\beta$, evaluated at various laser powers is plotted in Fig.~{S3}(b).

\begin{figure}[h]
     \centering
     \includegraphics[width=0.8\linewidth]{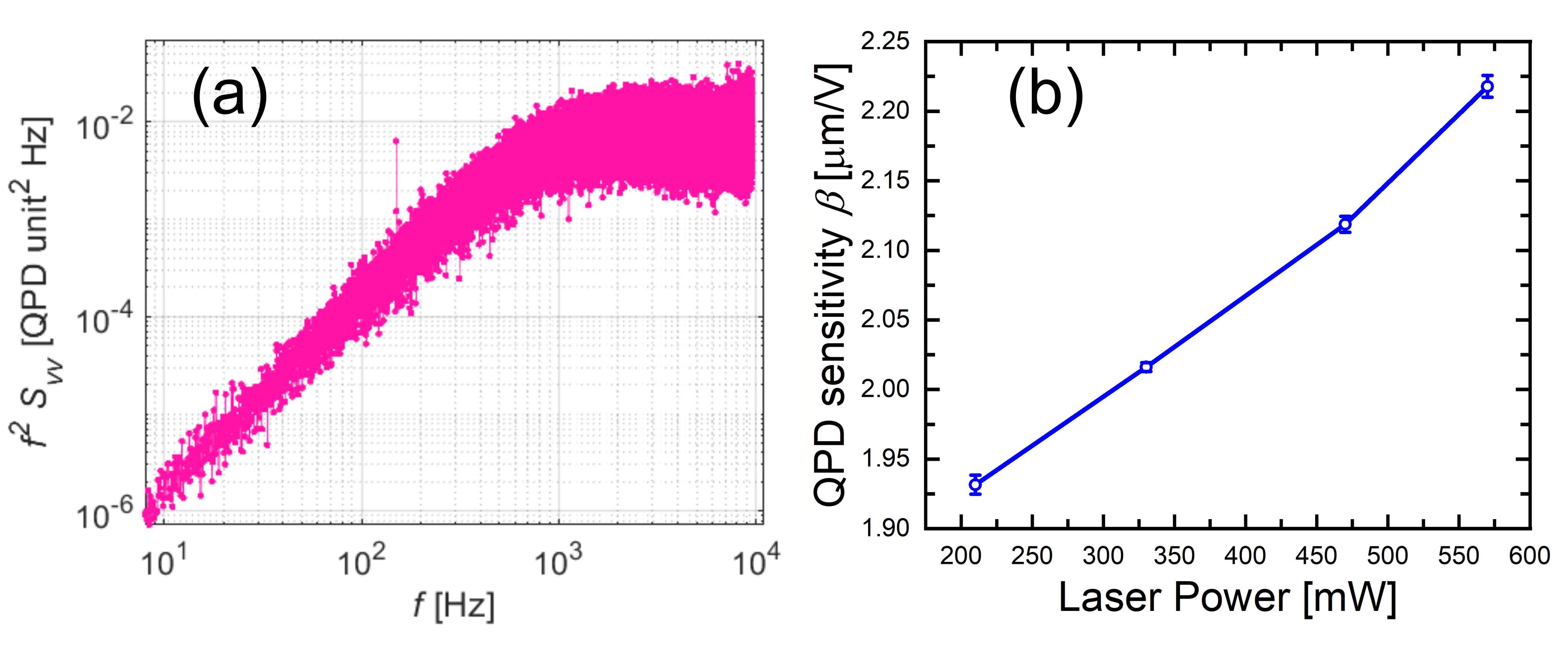}
     \caption{{(a)} $f^2S_{vv}$ as a function of $f$ for a microsphere trapped in water. {(b)} QPD sensitivity, $\beta$  \textit{vs.} laser power. The error bars represent the standard errors estimated from 11 microspheres trapped in water.}
     \label{fig:enter-label}
 \end{figure}
\newpage
 \subsection{Calculation of power spectral density (PSD) from the position fluctuations of a trapped Brownian microsphere}

 {Figure~\ref{fig:si3}(a) shows the thermal fluctuations of a trapped microsphere.} We acquired the time-dependent position fluctuations of an optically trapped microsphere for a period of 70 minutes. We divided this data into equally-spaced datasets containing position fluctuations acquired for 80 seconds. The dataset containing position fluctuations for 80 seconds was further divided into 10 sub-datasets with equally-spaced data. The power spectral density (PSD) for each sub-dataset was estimated using the periodogram function in MATLAB as shown by grey curves in Fig.~\ref{fig:si3}(b). 10 power spectral density curves were extracted to compute an average (black circles in Fig.~\ref{fig:si3}(b)). Since there can be noise and aliasing effects at low and high frequencies, power spectral densities {in} a frequency range of 10 to 11000 Hz were used for further analysis {(Figs.~\ref{fig:si3}(c,d))}. The peak observed near 200 Hz in {Figs.~\ref{fig:si3}(c,d)} was identified as environmental noise from the laboratory and eliminated. Despite carefully positioning the sample cell on the sample holder, a slight misalignment still {existed} between the direction of the applied field and the horizontal axis of the QPD. The slight misalignment resulted in a weak peak at $f = f_{AC}$ in the PSD along the $\hat{y}$-direction. We resolved this issue during data analysis by rotating the position fluctuation data until the electric field direction was perfectly aligned with the horizontal axis of the QPD. This reduces complexity in further analysis. 
 

 \begin{figure}[H]
     \centering
     \includegraphics[width=0.8\linewidth]{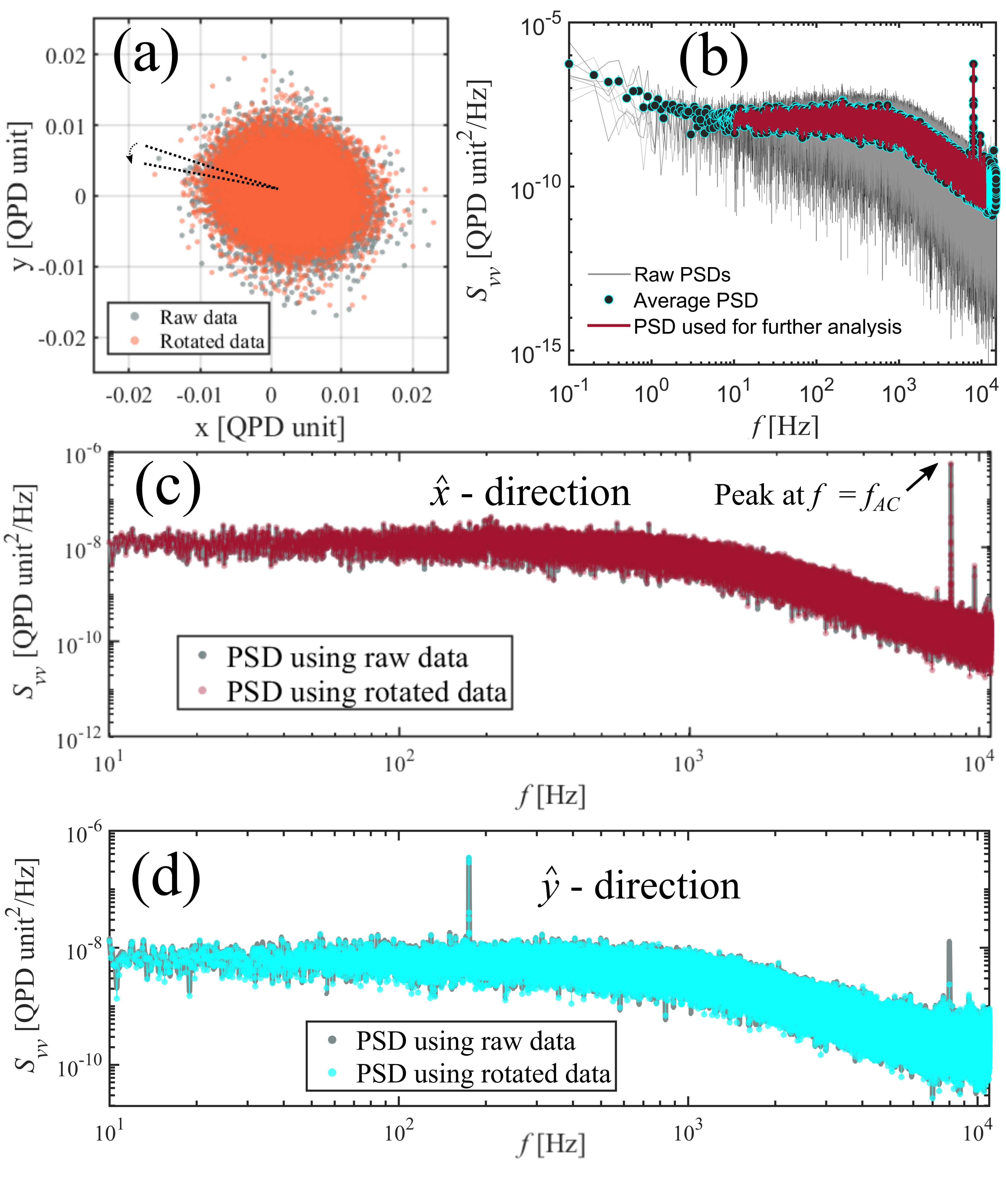}
     \caption{{(a)} Scatter plot showing position fluctuation{s} of the trapped microsphere. The data is rotated by an angle < 0.15 rad to align the electric field with the horizontal axis of the QPD. {(b)} Power spectral densities of individual sub-datasets, average power spectral density, and cropped power spectral density used for further analysis. {(c,d)} Power spectral densities along $\hat{x}$ and $\hat{y}$ directions respectively, computed from raw and rotated position fluctuations. It is seen from (d) that rotation eliminates the $f=f_{AC}$ peak from the $\hat{y}$-direction.}
     \label{fig:si3}
 \end{figure}

   \begin{figure}[h]
     \centering
     \includegraphics[width=1.0\linewidth]{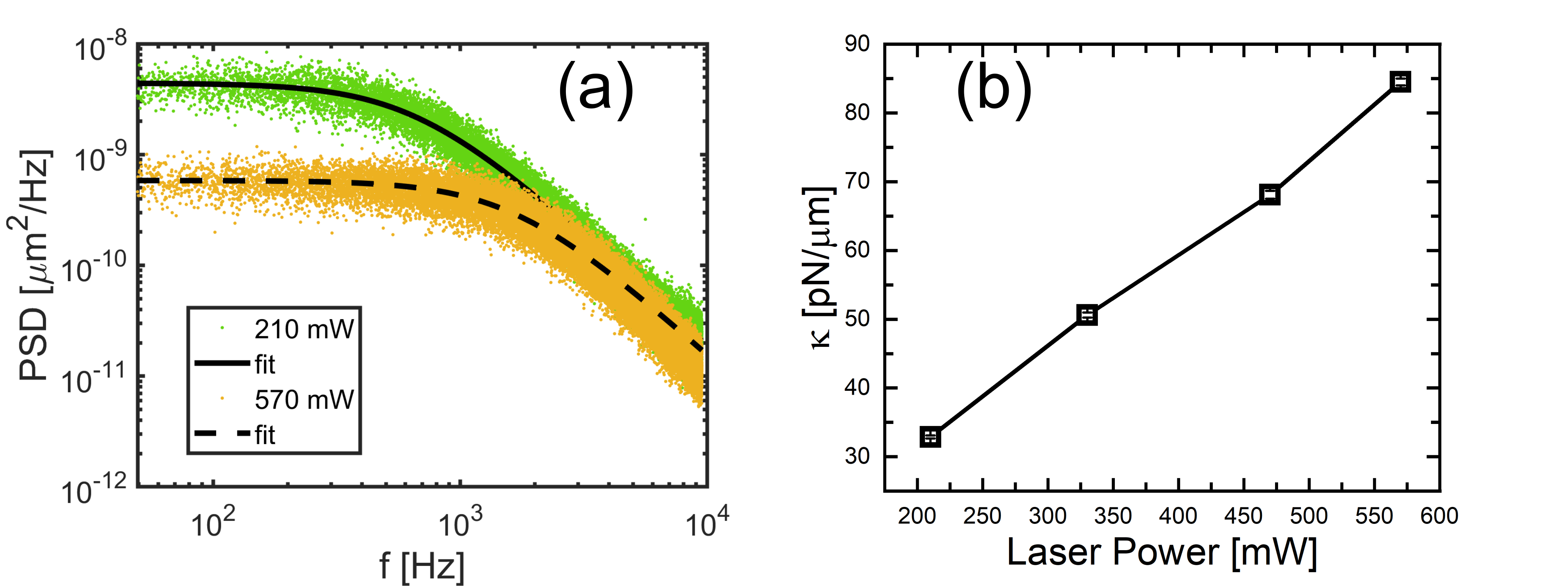}
     \caption{(a) PSDs for a trapped microsphere in water at two different laser powers and the Lorentzian fits to the data (black lines). (b) Changes in trap stiffness with laser power.}
     \label{fig:enter-label}
 \end{figure} 
\subsection{Calibration of trap stiffness}
 The stiffness of the optical trap was calculated using the power spectral density method. A Latex microsphere of size $1 {\mu}$m was trapped {in water} using an optical tweezer at a desired laser power. 
 The microsphere was trapped at approximately $  8 {\mu}$m above the bottom surface to eliminate wall effects. 
 The calculated power spectral density, $S_{xx}(f)$, of {the} centre of mass fluctuations {of the trapped particle} was fitted to a Lorentzian function:
 $${S_{xx}}=\frac{k_B T}{\pi^2\gamma (f_c^2+f^2)}$$
 {Data and f}its {for} two different laser powers are shown in Fig.~S5(a). The corner frequencies $f_c=k/2\pi\gamma$ were extracted from the fits. Using the fitted value of $f_c$ and the estimated value of {the Stokes drag coefficient} $\gamma = 6\pi \eta a$, the trap stiffness $\kappa$ was determined for four different laser powers. The trap stiffness shows a linear increase with laser power, as shown in Fig.~S5(b).     

 
\end{document}